\documentclass[printer]{aa} 
\usepackage{graphicx}
\usepackage[comma,authoryear]{natbib}
\usepackage{amsmath, amssymb} 
\usepackage{import}
\usepackage{xspace}
\usepackage{ulem}
\begin{document}

\newcommand\micron[0]{\,$\mu$m\xspace}
\newcommand\msun[0]{\,M$_{\odot}$\xspace}
\newcommand\e[1]{\,$\times$\,10$^{#1}$}
\newcommand{\hdzco}{H$^{13}$CO$^{+}$\xspace}
\newcommand{\dzco}{$^{13}$CO\xspace}
\newcommand\hii{H{\sc ii}\xspace}
\newcommand{\hi}{\hbox{H{\sc i}}\xspace}
\newcommand{\htwo}{\hbox{H$_{2}$}\xspace}

%

\title{Triggered/sequential star formation? A multi-phase ISM study around the prominent IRDC G18.93-0.03}

   \author{J. Tackenberg\inst{1},
  H. Beuther\inst{1},
  R. Plume\inst{2},
  T. Henning\inst{1},
  J. Stil\inst{2},
  M. Walmsley\inst{3},
  F. Schuller\inst{4},
  A. Schmiedeke\inst{5}
          }

\authorrunning{J. Tackenberg et al., 2012}
\titlerunning{Triggered/sequential star formation?}

          \institute{Max-Planck-Institut f\"ur Astronomie (MPIA), K\"onigstuhl 17, 69117 Heidelberg, Germany\\
            \email{last-name@mpia.de}
            \and
            Department of Physics \& Astronomy and the Institute for Space
            Imaging Science, University of Calgary, Calgary, AB T2N 1N4,
            Canada
            \and
            INAF-Osservatorio di Arcetri, Largo E. Fermi 5, 50125, Firenze,
            Italia
            \and
            European Southern Observatory, Alonso de Cordova 3107, Vitacura,
            Santiago, Chile
            \and
            I. Physikalisches Institut der Universit\"at zu K\"oln, Z\"ulpicher Strasse 77, 50937 K\"oln, Germany
          }
          
   \date{Received September 15, 1996; accepted March 16, 1997}

 
  \abstract
  {Triggered star formation has been discussed for many years, and evidence for the formation of stars and cores triggered by \hii regions is under debate.}
   {We investigate the imprints of an expanding \hii region on a pre-existing starless clumps.}
   {We selected an infrared dark filament spanning 0.8\degr.
     One portion of this filament, G18.93-0.03 is a prominent dust complex, with the
     molecular clump G18.93/m being IR dark from near IR wavelength up
     to 160\micron. Spitzer composite images show an IR bubble spatially
     associated with G18.93-0.03. We use
     GRS \dzco and IRAM 30m \hdzco data to disentangle the large and small
     scale spatial structure of the region. From ATLASGAL submm data we
     calculate the gas mass, while we use the \hdzco line width to estimate
     its virial mass. To study the IR properties of G18.93/m we use HERSCHEL
     data and produce temperature maps from fitting the spectral energy
     distribution. With the MAGPIS 20\,cm and SuperCOSMOS H$_{\alpha}$ data we
     trace the ionized gas, and the VGPS \hi survey provides information on
     the atomic hydrogen gas.}
   {We show that the bubble is spatially associated
     with G18.93, located at a kinematic near distance of 3.6\,kpc.
     The total gas mass of $\sim$ 870\msun splits up into 6 sub-clumps, of
     which G18.93/m is the most massive with 280\msun. The virial analysis
     shows that it may be gravitationally bound and has neither Spitzer young
     stellar objects nor mid-IR point sources within. Therefore we call it
     pre-stellar. Fitting the spectral energy distribution reveals a
         temperature distribution that decreases towards its center, but
     heating from the ionizing source puts it above the general ISM
     temperature. We find that the bubble is filled by \hii gas,
     ionized by an O8.5 star. Between the ionizing source and the IR dark
     clump G18.93/m we find a layered structure of hydrogen phases, from
     ionized to atomic to molecular gas, revealing a photon dominated
     region. Furthermore, we identify an additional velocity component within
     the bubble's 8\micron emission rim at the edge of the infrared dark cloud
     and speculate that it might be shock induced by the expanding \hii region.}
   {While the elevated temperature allows for the build-up of larger
     fragments, and the shock induced velocity component may lead to
     additional turbulent support, the density profile of G18.93/m does not show signatures of the expanding bubble. While the first two
     signatures favor high-mass star formation, we do not find conclusive
     evidence that the massive clump G18.93/m is prone to collapse because of
     the expanding \hii region.}


   \maketitle
%

\section{Introduction}
Based on the Galactic plane survey GLIMPSE \citep{Benjamin2003}, it has become possible to study bubble-like structures in the Milky Way in a
statistical sense. Initially, \citet{Churchwell2006} found a 'Bubbling Galactic Disk' and identified 322 bubbles. Today, several compilations of
bubbles in the Milky Way exist \citep[e.g ][]{Churchwell2006, Churchwell2007}.
The most extensive of which has been compiled by 'The Milky Way
Project' \citep{Simpson2012}, a citizen science project. They visually identified and classified more than 5000 bubbles. Infrared bubbles are defined
by their bright 8\micron emission rims. Although their origin may be manifold, many of the bubbles have shown to be created by a, not necessarily
central, ionizing source. In this context, the emission rim originates from UV excited polycyclic aromatic hydrocarbons (PAHs). While all hydrogen
ionizing photons are absorbed within the borders of an ionization front, lower
energy UV photons pass through the ionization front and may excite the
PAHs. In addition, the ionizing source heats its surrounding gas, which expands and drives a shock front beyond the ionization front. In
between the shock and ionization front, neutral material can be accumulated. \citet{Elmegreen1977} discussed the triggering of star formation on the
borders of expanding H{\sc ii} regions. Among others, 'collect and collapse' describes the collapse of swept up material along a compressed
layer. \citet{Deharveng2003} and \citet{Zavagno2006} established 'collect and collapse' observationally. A brief overview on triggered star formation is given in \citet{Deharveng2010}.

Other results of the GLIMPSE survey are systematical and sensitive
  studies of infrared dark clouds \citep[IRDCs, e.g.][]{Peretto2009}. Previously
  discovered by the two infrared satellites ISO and MSX, their high column
  densities absorb the background emission even at infrared wavelengths and thus become visible as dark patches. With such high column densities, IRDCs are believed
to be the cradles of the next generation of stars \citep{Rathborne2006,
  Simon2006}. In addition, a rising number of large scale Galactic Plane surveys allow one to trace the various stages of star formation.
From mm and sub-mm observations of cold, dense clumps and cores
\citep{Schuller2009, Bally2010} to mid- to near-IR observations of young
stellar objects \citep{Benjamin2003, Carey2009, Molinari2010} and stars as well as the ionized and molecular gas \citep{Condon1998, Jackson2006, Purcell2010}, an almost complete picture can be obtained for regions within the Galactic Plane. 

\begin{figure*}[tbp]
  \includegraphics[width=1.0\textwidth]{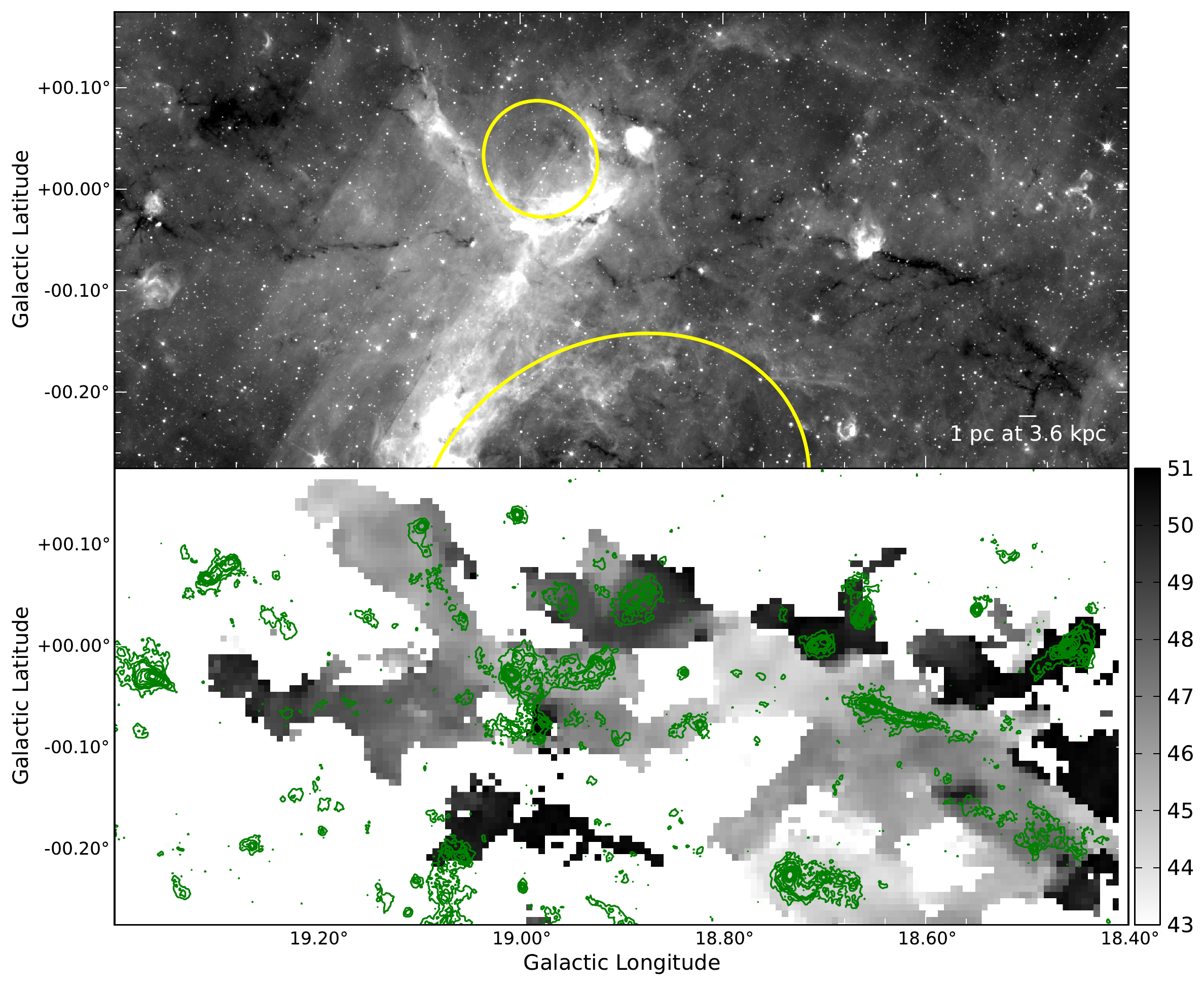}
  \caption{An IRDC filament along the Galactic plane, spanning
    across from the lower right (18.45\degr,-0.20\degr) to the upper left
    (19.30\degr, -0.02\degr). The top panel shows a 8\micron SPITZER image and
    the color stretch is chosen to bring the IRDC to prominence. The lower
    panel presents the intensity weighted peak velocity (moment 1) map of the
    GRS \dzco data. Regions with velocities outside the 43\,km/s to 51\,km/s are
    clipped. The green contours are from ATLASGAL with contour levels at
    0.15\,Jy or 3\,$\sigma$,
    0.3\,Jy, 0.4\,Jy, 0.5\,Jy, 0.7\,Jy, 0.9\,Jy,
    1.3\,Jy, 1.8\,Jy, and 2.5\,Jy. The yellow ellipses indicate the
      bubbles as given in \citet{Simpson2012}.}
  \label{fig:glimpse_filament}
\end{figure*}
However, although both filamentary structures and IRDCs have gotten much
interest over the last years, extreme IRDCs like the 80\,pc long ``Nessie''
\citep{Jackson2010} are still rare objects. A similar large scale
  filament of more than 50\,pc length has been presented by
\citet{Kainulainen2011}. Figure \ref{fig:glimpse_filament} shows this filament,
spanning over $\sim$\,0.87\degr. At a kinematic near distance of $\sim$
  3.6$^{+1.0}_{-0.5}$\,kpc, this translates to a length of $\sim$
54\,pc. (For a discussion of the distance, see
Sec. \ref{sec:velo_discussion}.) Its coherent velocity structure is shown in
the bottom panel of Fig. \ref{fig:glimpse_filament}. Following the low density
gas traced by \dzco along the extinction patch, the smooth velocity
transitions suggest that the entire region is spatially
connected. \citet{Kainulainen2011} studied the extinction map of the region
and estimate the total gas reservoir to be 4.7\e{4}\msun. Along the filament
 at lower longitudes, IRDC\,18223 is a well-studied prototype IRDC with ongoing high-mass star formation (e.g. \citealt{Beuther2010} and references therein). 

A particularly interesting region in the context of triggered star formation
is the prominent IRDC G18.93. Being part of the large scale filament
  described above, its location is also coincident with the rim of a IR bubble
just above the filament (indicated by the upper yellow ellipse in
Fig. \ref{fig:glimpse_filament}). We will show that its dense gas, the
filament and the bubble have coherent velocities and therefore are all
at the same distance. Using CLUMPFIND on 870\micron dust continuum data we identify six
sub-clumps. The most massive clump, denoted G18.93/m, is even infrared dark from near infrared wavelengths up to 160\micron.

For the particular case of G18.93/m we will address the following questions:
Do we see evidence that the bubble influences G18.93/m? Do the properties of G18.93/m differ from other high-mass starless clumps?

We will first introduce the data used for the analysis and their
implications. In Sec. \ref{sec:results} we introduce G18.93 and the
neighboring bubble, and supplement a more detailed description in
  Sec. \ref{sec:details_G1893} and onwards. Sec. \ref{sec:imprints} describes
the signs of triggered star formation we find and discusses their consequences. We summarize the paper in a concluding section. 


\section{Description and implications of observations and data}
We used various publicly available surveys, as well as IRAM 30\,m follow-up observations of the specific region. 

\subsection{The cold dust tracing the molecular hydrogen, ATLASGAL}
\label{ATLASGAL}
Molecular hydrogen, H$_2$, is hard to trace directly, and no pure
rotational transitions are observable from the ground. 
However, in regions of high particle densities the gas thermally couples with
the dust.
Cold dust can be observed by its
thermal radiation in the Rayleigh-Jeans regime at submm and mm wavelength. The
APEX Telescope Large Area Survey of the Galaxy (ATLASGAL,
\citealt{Schuller2009}) covers the full inner Galactic plane at 870\,$\mu$m
with a resolution of 19.2\arcsec and an rms below 50\,mJy. 
Using the Large APEX Bolometer CAmera (LABOCA, \citealt{Siringo2009}),
each position of the inner Galactic plane has been mapped on-the-fly twice,
each in different scanning directions. The pointing is on the order of
4\arcsec and the calibration uncertainty is lower than 15\%. A detailed
description of the data reduction is given in \citet{Schuller2009}.
The submm emission is mostly optically thin.

\subsection{IRAM 30\,m observations}
We used the IRAM 30\,m telescope to map the region around G18.93 identified in the ATLASGAL
870\,$\mu$m images \citep{Schuller2009, Tackenberg2012} in H$^{13}$CO$^{+}$  and
SiO. 

H$^{13}$CO$^{+}$ is a well established and common dense gas tracer with a critical density at 20\,K of $\sim$\,1.8\e{5}\,cm$^{-3}$. It is
generally optically thin and yields information on the bulk motion of the gas
\citep{Vasyunina2011}. SiO becomes released from the dust grains due to shocks 
and is therefore a well known tracer of shocked gas and molecular outflows \citep[e.g. ][]{Schilke1997}.

During our observing run in August 2010 the weather conditions were poor with average precipitable water vapour (pwv) $>$ 10\,mm. Although
the weather affected the pointing, it is, nonetheless, better than a third
of the 3 mm beam (or $\sim$ 9\arcsec). We used on-the-fly mapping on 5 boxes
to cover all 870\,$\mu$m continuum emission above 0.3\,Jy, or
2\,$\times$\,10$^{22}$\,cm$^{-2}$ (assuming a general dust temperature of
15\,K, for further details see Sec. \ref{sec:dust_mass}; see
  Fig. \ref{fig:glimpse_general} for the coverage of the observations).
The IRAM 30\,m beam width for both H$^{13}$CO$^{+}$ and
SiO is 29.9\arcsec and the 3\,mm setup chosen provides a spectral resolution of 40\,kHz at
40\,MHz bandwidth. This translates to a native velocity resolution of
0.14\,km/s at 86\,GHz. The data reduction was done using CLASS from the
GILDAS\footnote{http://www.iram.fr/IRAMFR/GILDAS} package. We subtracted a
linear baseline and removed the three central channels to avoid occasional
spikes from the backends. The maps have been produced with a pixel scale of 14.2\arcsec. With a total integration time of 11.3\,h the average rms at a velocity resolution of 0.25\,km/s is 0.08\,K and 0.05\,K for
H$^{13}$CO$^{+}$ and SiO, respectively.

Simultaneously with the 3\,mm observations, we recorded two of the H$_2$CO(3-2)
lines, CH$_3$OH(4-3), and HC$_3$N(24-23) at 1\,mm. However, due to the poor
weather conditions the average rms noise in the spectra is above 0.32\,K. With
the given sensitivity, for the region of interest we do not detect any of the
higher excitation lines at 1\,mm.



\subsection{The large scale cloud complex in H{\sc i}, VGPS}
The ISM is composed primarily of atomic hydrogen. H{\sc i} (self) absorption features are usually spatially
correlated to molecular clouds and IRDCs \citep{Li2003}. We used the 21\,cm H{\sc i} observations of
the VLA Galactic Plane Survey (VGPS, \citealt{Stil2006}) to study the large
scale distribution of atomic hydrogen. With the VLA in D configuration,
  the best resolution at 21\,cm is 45\arcsec. The interferometer data were
  supplemented with short spacings from the NRAO Greenbank telescope (GBT).
To improve the survey's sensitivity to extended, low surface bright emission, the data were reduced to a synthesized beam of 
1\arcmin at a velocity resolution of 1.56\,km/s. The data cubes are sampled to
a pixel scale of 18\arcsec at 0.8\,km/s. The calibration was made to be
  consistent with the NVSS survey \citep{Condon1998}.


\subsection{A large-scale view of the molecular cloud, GRS}
\label{GRS}
To study the molecular cloud component to its full extent we selected $^{13}$CO, a
(well known) low density gas tracer with a critical density
$\sim$\,1.9\e{3}\,cm$^{-3}$. While it traces the same density range as
$^{12}$CO, the isotopologue $^{13}$CO has lower abundances and is therefore
optically thinner. The Boston University-FCRAO Galactic Ring Survey (GRS,
\citealt{Jackson2006}) provides spectroscopic $^{13}$CO (1-0) observations of
the Galactic plane for Galactic longitudes 18$^{\circ}$ $\le$ l $\le$
58$^{\circ}$. The observations were performed with the 
single sideband focal plane receiver SEQUOIA, mounted on the Five College
Radio Astronomy Observatory (FCRAO) 14\,m telescope in OTF mode.
Its angular resolution is 46\arcsec, and the publicly available maps
are sampled to a grid of 22\arcsec. The velocity resolution is 0.2\,km/s,
the pointing accuracy is better than 5\arcsec.

\subsection{The stellar component of the complex, UKIDSS, GLIMPSE and MIPSGAL}
To examine the stellar content and stellar properties, we employed near- and
mid-IR surveys. Both the Galactic Legacy Infrared Mid-Plane Survey
Extraordinaire (GLIMPSE, \citealt{Benjamin2003}) and
the MIPS Galactic plane survey (MIPSGAL, \citealt{Carey2009}) are SPITZER
Legacy Programs. The GLIMPSE survey provides maps of the inner Galactic plane
for -60$^{\circ}$ $\le$ l $\le$ 60$^{\circ}$ at 3.8\,$\mu$m, 4.5\,$\mu$m,
5.8\,$\mu$m, and 8.0\,$\mu$m. The resolution is 1.7\arcsec, 1.7\arcsec,
1.9\arcsec, 2.0\arcsec, respectively. From MIPSGAL we used the 24\,$\mu$m
images with a resolution of 6\arcsec. Additional near-IR JHK data have been
taken from the UKIDSSDR7PLUS data release \citep{Lawrence2007}. 

\subsection{The peak of the SED, HiGal}
\label{sec:higal_description}
Covering the peak emission of the spectral energy distribution (SED) of cold
dust, the HIGAL/HERSCHEL survey \citep{Molinari2010} provides mid-IR
observations at 70\,$\mu$m and 160\,$\mu$m, and at submm wavelength
250\,$\mu$m, 350\,$\mu$m, and 500\,$\mu$m. Therefore, they connect the ATLASGAL
submm observations of the cold dust at 870\,$\mu$m to the near- and mid-IR
observations of the hot dust and stellar radiation. From a black body fit to
the fluxes, we will estimate the temperature of the dust. The
observations were carried out in parallel mode with a scanning speed
of 60\arcsec/s. Since the official, fully reduced HiGal fields are
not yet available, we reduced the HiGal raw data using HIPE
(level 0 to 1, \citealt{Ott2010}) and SCANAMORPHOS (level 1 to 2, v16.0,
\citealt{Roussel2012}). The angular
resolutions at the given wavelengths are 10.2\arcsec, 13.6\arcsec,
23.4\arcsec, 30.3\arcsec, and 42.5\arcsec, respectively \citep{Traficante2011}.

\subsection{The ionized gas, MAGPIS and SHS}
\label{sec:intro_ionized}
In addition to thermal dust and line emission, free-free emission can be
observed, which traces ionized gas. 
A certain amount of ionizing photons per second is required to maintain a
given amount of gas in ionized state. This allows a characterization of the
ionizing source. However, especially at cm wavelength another significant
  emission contribution is coming from the cosmic ray electrons' synchrotron radiation. Here, the spectral index can help to distinguish which contribution dominates.

The Multi-Array Galactic Plane Imaging Survey (MAGPIS, \citealt{Helfand2006})
mapped the 1.4 GHz, or 20 cm, continuum of the first Galactic quadrant
partially. Using
  the VLA in B, C, and D configurations allows a resolution of $\sim$
  6\arcsec, while the extended structures are preserved by including
  Effelsberg 100\,m data \citep{Reich1990}. The ready reduced maps have a pixel scale of
  2\arcsec/pix.  

While being very susceptible to extinction, the optical H$_{\alpha}$ line traces the down cascade of an electron after recombining from ionized to atomic hydrogen. The
SuperCOSMOS H$_{\alpha}$ survey \citep{Parker2005} digitized AAO/UKST narrow-band H$_{\alpha}$ observations of the southern Galactic Plane with a resolution
of $\sim$ 1\arcsec.

\section{Results}
\label{sec:results}
The employment of the various surveys allows us to draw a comprehensive
large scale picture of the cloud complex associated with 
the IRDC G18.93. In the following we will describe the environment, going from large to small scales. Finally we will use velocity information to study the spatial relations.

\subsection{A massive IRDC in the vicinity of a bubble}
\label{sec:irdc_bubble}

As mentioned in the introduction, a prominent feature along the $>$\,50\,kpc
long filament is the bubble like structure indicated by the yellow ellipse in
Fig. \ref{fig:glimpse_filament}. A second bubble, N24 from
\citet{Churchwell2006, Deharveng2010} marked by the lower yellow ellipse in
Fig. \ref{fig:glimpse_filament}, is close in
projection as well. However, while the filament has a velocity
  (v$_{lsr}$) around $\sim$\,45\,km/s, the second bubble has a distinct
  velocity at around 60\,km/s and therefore is at a different distance.

\begin{figure*}[tbp]
  \includegraphics[width=1.0\textwidth]{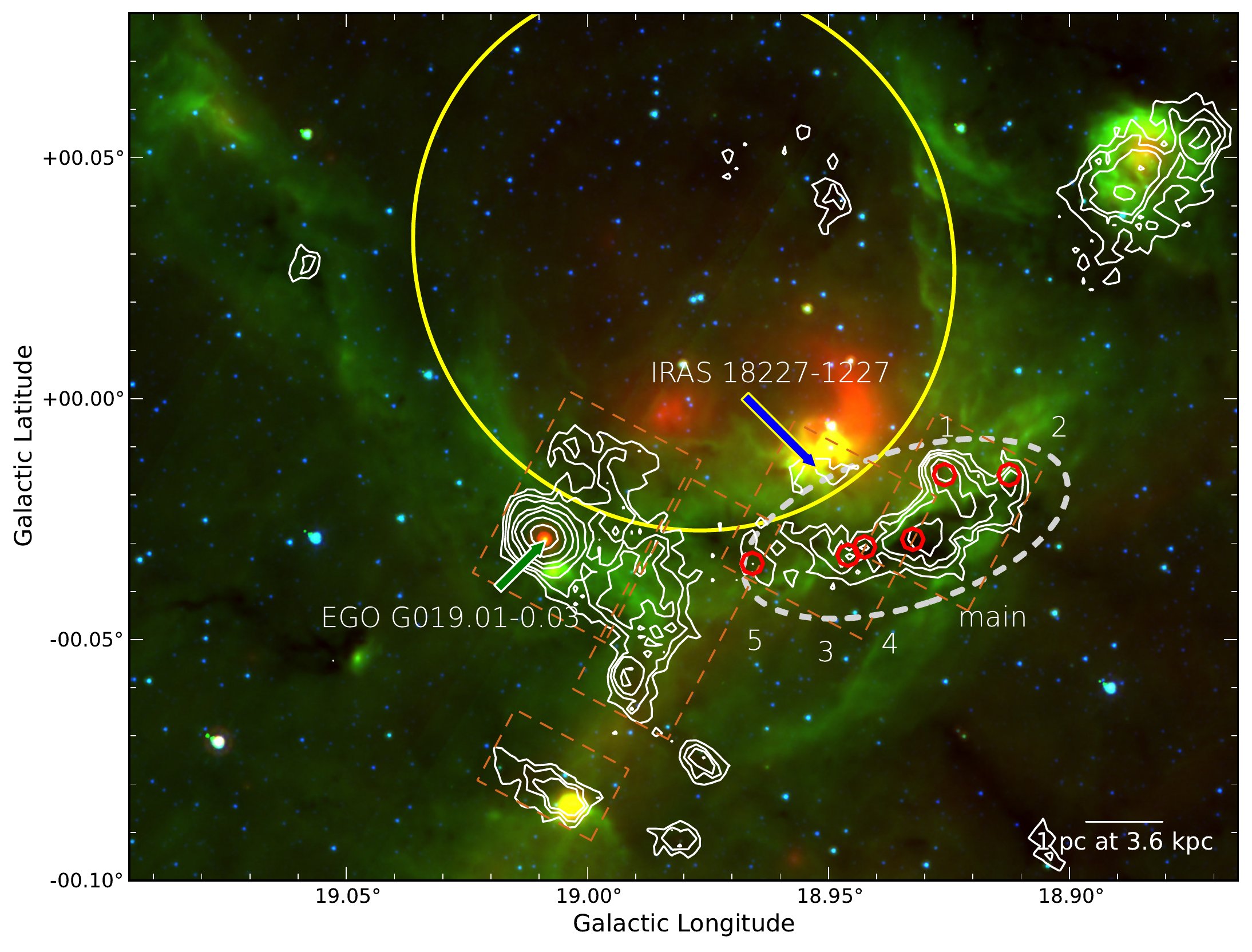}
  \caption{Three color image of the bubble connected to the long filament, MWP1G018980+00304, composed from the SPITZER 3.6\micron, 8\micron, and 24\micron bands in blue, green, and red,
    respectively. White contours are from ATLASGAL (at 0.3\,Jy, 0.4\,Jy, 0.5\,Jy, 0.7\,Jy, 0.9\,Jy,
    1.3\,Jy, 1.8\,Jy, and 2.5\,Jy) for which the numbers denote the CLUMPFIND peaks, marked by the red circles. The yellow solid ellipse marks the bubble with interpolated bubble dimensions adopted from \citet{Simpson2012}. The two arrows mark
    IRAS 18227-1227, and the massive proto-stellar object EGO
    G019.01-0.03. While the dashed orange boxes indicate the areas
    that have been mapped in \hdzco with the IRAM 30\,m telescope, the gray dashed ellipse is drawn around the ATLASGAL contours connected to G18.93.}
  \label{fig:glimpse_general}
\end{figure*}

A zoom in on the smaller bubble closer to the filament is shown in Fig. \ref{fig:glimpse_general}. 
It opens towards higher Galactic latitudes and only little 24\micron emission is
visible within the bubble. The opening of the bubble resembles a typical 'champagne flow'
\citep{Stahler2005} with lower density material at higher
latitudes.
On the lower inner edge of the rim a bright IRAS source is located together with two bright NIR sources.


Directly at the edge of the bubble at lower latitudes, several studies
identified infrared dark clouds (IRDCs) and massive dense clumps,
e.g. SDC G18.928-0.031 \citep{Peretto2009}, [SMC2009] G18.93-0.01
\citep{Schuller2009}, or BGPS G18.926-0.019 \citep{Rosolowsky2010}. As
shown in Fig. \ref{fig:glimpse_general}, the ATLASGAL survey at
870\micron revealed several dense structures along the bubble. The
largest structure in Fig. \ref{fig:glimpse_general} is partly
following the IRDC filament. In addition, \citet{Cyganowski2008}
found an extended green object (EGO) within the rim of the bubble,
EGO\,G19.01-0.03 which drives a bipolar outflow and is characterized
as a genuine fast accreting massive young stellar object
\citep{Cyganowski2008, Cyganowski2011, Cyganowski2011a}.
\begin{figure*}[tbp]
  \includegraphics[width=\textwidth]{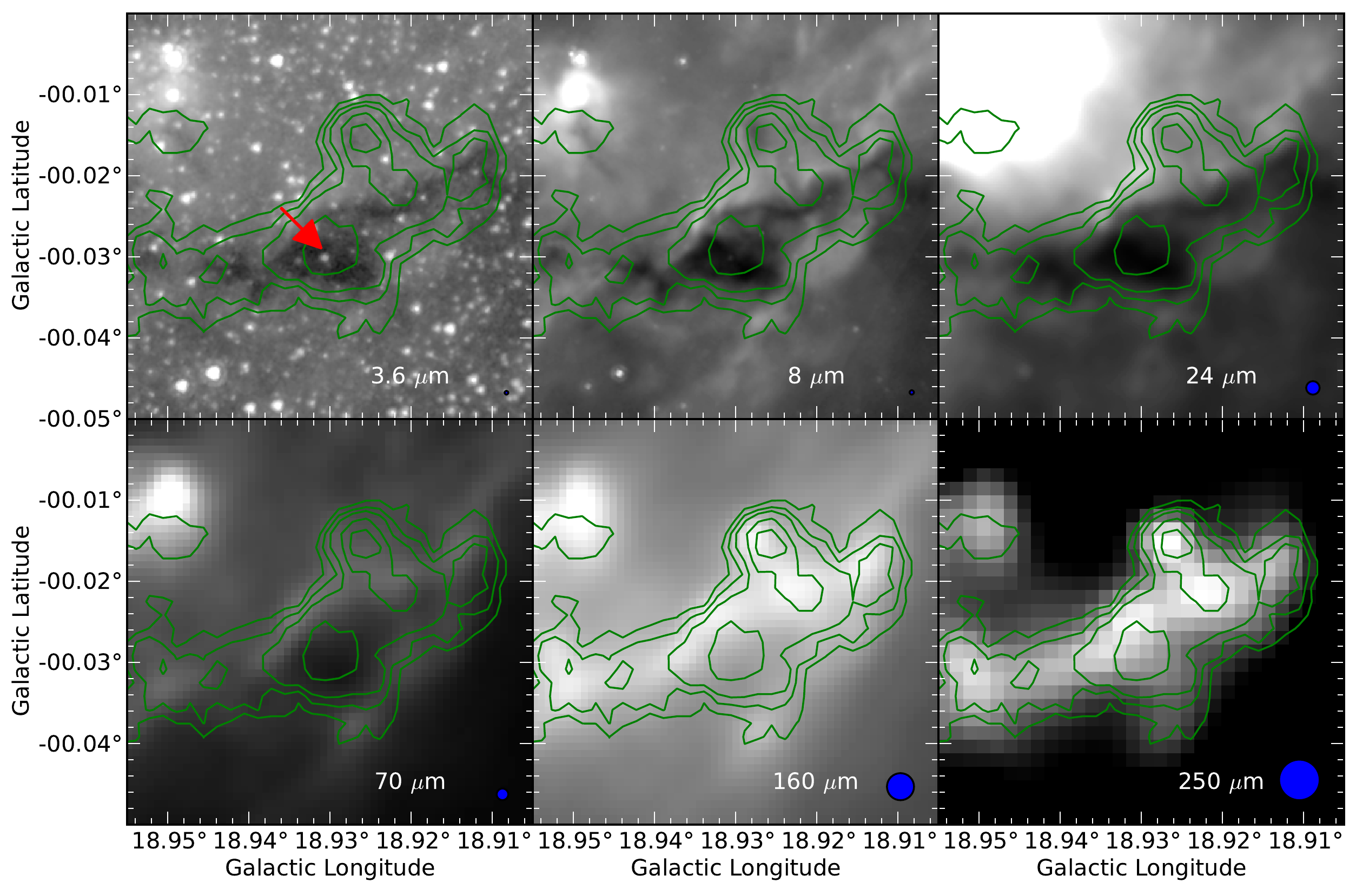}
  \caption{A multi-wavelength zoom on G18.93/m, the continuum peak connected
    to the prominent IR extinction feature. The panels are (from left to right
    in reading order): 3.6\micron, 8.0\micron, 24\micron, 70\micron,
    160\micron, 250\micron, from GLIMPSE, MIPSGAL, and HIGAL. The blue circles
    in the bottom right of each panel indicate the beam sizes, the green
    contours are from ATLASGAL (at 0.3\,Jy, 0.4\,Jy, 0.5\,Jy, 0.7\,Jy, 0.9\,Jy,
    1.3\,Jy, 1.8\,Jy, and 2.5\,Jy). The colors of the source marked by a red
    arrow in the first
    panel are not consistent with typical colors of young stellar objects (class 0/1) \citep{Gutermuth2008, Robitaille2008}.}
  \label{fig:G1893_zoom}
\end{figure*}
As shown in Fig. \ref{fig:G1893_zoom}, the main emission peak of the
IRDC shows up in absorption up to even 160\micron. While absorption of
  IRDCs usually refers to extinction against the Galactic background from PAHs
  and very small grains, it is unclear whether the absorption feature at the
  longer wavelength of 160\micron implies a lack of emission or simply no
  grains hot enough to emit in that line of sight. 
However, the part around EGO G019.01-0.03 and all connected continuum
emission towards lower galactic latitudes does not even show 24\,$\mu$m
extinction signatures. 


Within and along the bubble, there are few additional complexes dense enough
to contain dense clumps and become visible at ATLASGAL sensitivity. However, since we want to study the influence of an expanding
\hii region upon an already existing filament, we will not discuss the
continuum emission in the region that is clearly above the filament. Due
to its characteristic absorption we will call the IR dark structure defined by
the ATLASGAL emission G18.93 (see gray ellipse on
Fig. \ref{fig:glimpse_general}). For reasons we will discuss in
Sec. \ref{sec:velo_discussion} we keep the ATLASGAL emission aligned with the
galactic latitude axis connected to EGO G019.01-0.03 separate.

It is interesting to note that both the massive young stellar object
EGO G019.01-0.03 and the IRDC filament are at the interface of the
bubbles shown in Fig. \ref{fig:glimpse_filament}. 

\subsection{Disentangling the spatial relations of G18.93 and EGO G19.01}
\label{sec:velo_discussion}
\begin{figure*}[tbp]
  \includegraphics[width=1.0\textwidth]{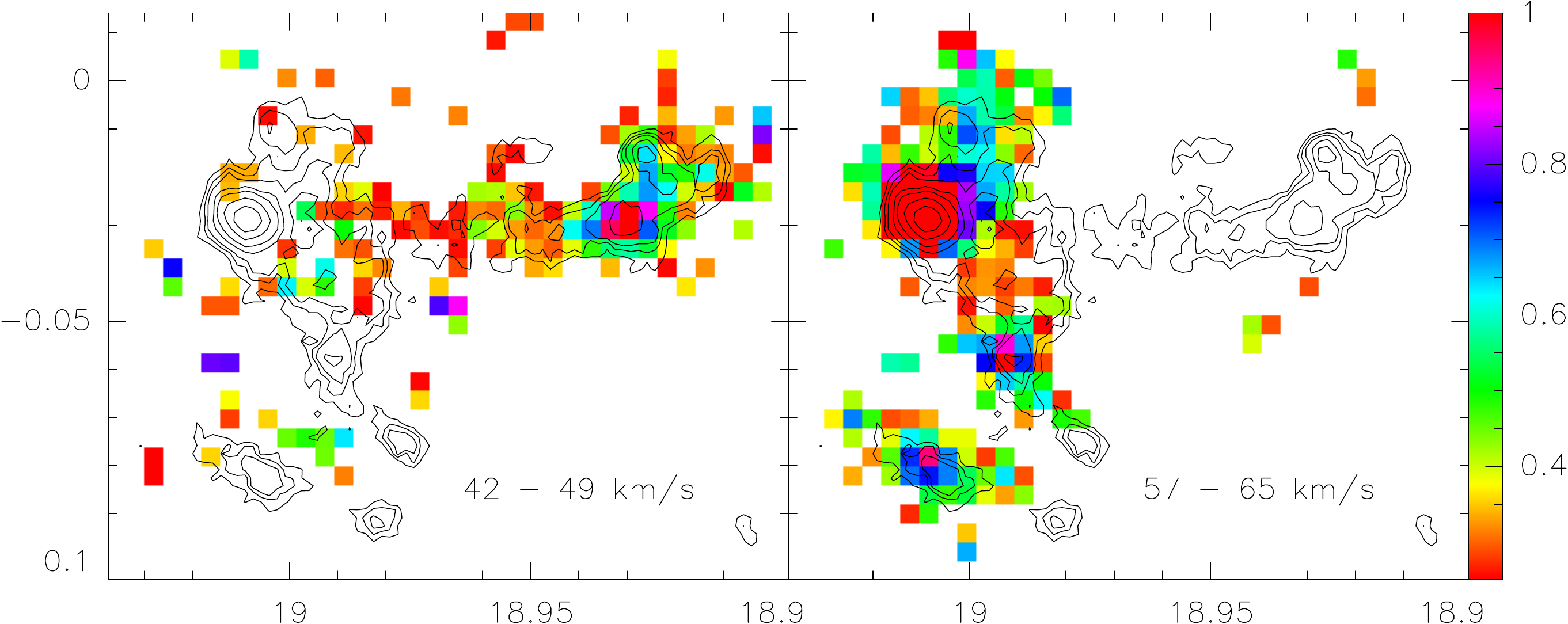}
  \caption{Color coded map of the H$^{13}$CO$^{+}$ emission in [K km/s] for the velocity
    regime 42\,km/s to 49\,km/s on the left, and 56\,km/s to 65\,km/s on the
    right. The black contours are from ATLASGAL with levels at 0.3\,Jy, 0.4\,Jy, 0.5\,Jy, 0.7\,Jy, 0.9\,Jy, 1.3\,Jy, 1.8\,Jy, and 2.5\,Jy.}
  \label{fig:H13COplus_general}
\end{figure*}
From \dzco, partly shown in Fig. \ref{fig:glimpse_filament}, it cannot
be concluded whether the dense dust for the region of interest seen in
ATLASGAL is part of the 45\,km/s filament component or the 60\,km/s
  component connected to the second bubble. To address that question and to disentangle the velocity structure of the dense molecular gas, we mapped G18.93 and the dust emission around EGO G19.01-0.03 in H$^{13}$CO$^{+}$, a dense gas tracer. 

As shown in Fig. \ref{fig:H13COplus_general}, the dense gas has two distinct velocity components. Consistent with earlier single spectra
\citep{Cyganowski2011, Wienen2012}, the IRDC G18.93 shows emission at velocities between 43\,km/s and 47\,km/s, while the peak of
EGO G19.01-0.03 and the ATLASGAL emission directly connected to it have velocities between 59\,km/s and 63\,km/s. For both components, the dense
molecular gas tracer resembles the structure defined by the ATLASGAL emission contours very well. Therefore, one can attribute the same velocities
unambiguously to the ATLASGAL emission. That is not obvious from the lower density $^{13}$CO alone. In addition, it is obvious that G18.93 is part of
the long filament and the primary bubble which is connected to it, while EGO G19.01-0.03, a site of
massive star formation and its connected material below this region, is
at velocities consistent with the larger bubble from \citet{Churchwell2006}. 

In order to constrain the distance to the filament and the connected bubble, we employ
the Galactic rotation curve given in \citet{Reid2009}. Assuming a
common velocity of 45\,km/s, the kinematic near distance becomes 3.6
kpc. From both extinction \citep{Kainulainen2011} and considerations
of the H{\sc i} absorption feature (see Sec. \ref{sec:pdr} for
details) we can conclude that the near distance is appropriate. Note
that \citet{Kainulainen2011} give a kinematic distance of 4.1\,kpc and
even larger distances from velocity independent methods. Therefore,
error propagation resulting in a distance uncertainty of 0.05\,kpc
underestimates the uncertainty. We expect the uncertainty rather to be on the
order of 0.5\,kpc. 


\subsection{Details and aspects of the G18.93 complex}
\label{sec:details_G1893} 
So far we have shown that there is a large filament along the Galactic plane that has a consistent velocity structure over more than 50\,pc. At
l\,=\,18.93\degr, b\,=\,0.03\degr the filament becomes very opaque, shows high mid-IR extinction and large gas column densities. 
Part of it shows IR absorption up to
160\micron. Just above G18.93, a bubble is visible. Its PAH rim visible at
8\micron has a direct interface with G18.93. We will show later that the
bubble is an expanding \hii region. Therefore, G18.93 is an ideal source in which
to
study the influence of massive stars on starless clumps and search for imprints of triggering.

\subsubsection{The dust component of G18.93}
\label{sec:dust_mass}
As mentioned in Sec. \ref{sec:velo_discussion} and shown in Fig. \ref{fig:H13COplus_general}, the densest gas (and most of the molecular hydrogen) in the complex is concentrated in the regions seen by ATLASGAL at 870\micron (see Sec. \ref{ATLASGAL}). 

In order to identify column density peak positions and associate masses to
individual clumps, we used the CLUMPFIND algorithm of
\citet{Williams1994}. Starting at a 6\,$\sigma$ contour of 0.3\,Jy, we chose
0.4, 0.5, 0.7, 0.9, 1.3, 1.8, and 2.5\,Jy as extraction thresholds. (For a detailed discussion of the extraction thresholds see \citealt{Tackenberg2012}.) The positions of the extracted clumps are listed in Table \ref{tab:clump_mass} and plotted in Fig. \ref{fig:glimpse_general}. We denote the prominent absorption dip and most massive clump G18.93/m.
\begin{table*}[htb]
  \caption{CLUMPFIND decomposition of dust clumps extracted on ATLASGAL 870\micron map. }
  \begin{tabular}{| p{2.cm} *{4}{l} *{1}{p{1.5cm}} *{1}{p{1.8cm}} *{3}{p{1.cm}} |} 
    \hline
    Sub-clump name & Gal. lon. & Gal. lat. & RA(2000) & DEC(2000) & HiGal peak temp & Peak column density & Mass & Angular radius & Radius \\
    & [ $^{\circ}$ ] & [ $^{\circ}$ ] & [ hh:mm:ss.s ] & [ dd:mm:ss ] & [ K ] &[ 10$^{22}$ cm$^{-2}$ ] & [ \msun ] & [ \arcsec ] & [ pc ] \\
    \hline
1      &  18.9259   &  -0.0158  &     18:25:32.4 & -12:26:47   &     23.0   &  4.2    &    248        &   27     &       0.46\\
main   &  18.9325   &  -0.0292  &     18:25:36.0 & -12:26:48   &     21.7   &  3.5    &    276        &   29     &       0.50\\
2      &  18.9125   &  -0.0158  &     18:25:30.9 & -12:27:29   &     22.3   &  2.4    &    104        &   21     &       0.36\\
3      &  18.9459   &  -0.0325  &     18:25:38.3 & -12:26:11   &     22.5   &  2.2    &    107        &   22     &       0.38\\
4      &  18.9425   &  -0.0308  &     18:25:37.6 & -12:26:19   &     22.4   &  2.1    &     57        &   15     &       0.27\\
5      &  18.9658   &  -0.0342  &     18:25:41.0 & -12:25:10   &     23.0   &  1.5    &     43        &   15     &       0.26\\
    \hline
  \end{tabular}
  \label{tab:clump_mass}
\end{table*}

Assuming optically thin thermal emission, the H$_2$ gas column density can be calculated from the dust continuum emission via 
\begin{equation}
  N_{gas} = \frac{ R F_{\lambda} }{ B_{\lambda}(\lambda,T) m_{H_2} \kappa \Omega}{\text ,}
\end{equation}
with the gas-to-dust ratio R\,=\,100, F$_{\lambda}$ the flux at the given wavelength, B$_{\lambda}$($\lambda$,T) the blackbody radiation as a function of wavelength and
temperature, m$_{H_2}$ the mass of a hydrogen molecule, and the beam size $\Omega$. 
Assuming typical beam averaged volume densities in dense clumps of 10$^5$\,cm$^{-3}$ and dust grains with thin ice mantles, the dust mass absorption
coefficient from \citet{Ossenkopf1994} becomes $\kappa$ = 1.42\,cm$^2$ g$^{-1}$ at 870\micron. The temperatures are taken from the HiGal temperature map described in Sec. \ref{sec:dust_temp}. Because of the sparse resolution, we adopted the temperature at each peak position. The peak column densities at the spatial resolution of ATLASGAL (19.2\arcsec) are then between 1.0\e{22}\,cm$^{-2}$ to 4.2\,$\times$\,10$^{22}$\,cm$^{-2}$.

With the distance as additional parameter and the integrated flux of the clump, the mass can be calculated in a similar way as given above, \begin{equation}
  M_{gas} = \frac{  R d^2 F_{\lambda, tot} }{ B_{\lambda}(\lambda,T) \kappa} \text{.}
  \label{eqn:mass}
\end{equation}
Assuming the kinematic distance of 3.6\,kpc, the total mass becomes $\sim$ 870\msun, with clump masses between $\sim$ 30\msun and 250\msun. Individual
column densities and masses are listed in Table \ref{tab:clump_mass}. The given radius is the effective radius for equating the pixel area of each
clump with a theoretical circular area, as calculated by CLUMPFIND.

\subsubsection{Temperature map of G18.93}
\label{sec:dust_temp}
\begin{figure}[tbp]
  \includegraphics[width=0.5\textwidth]{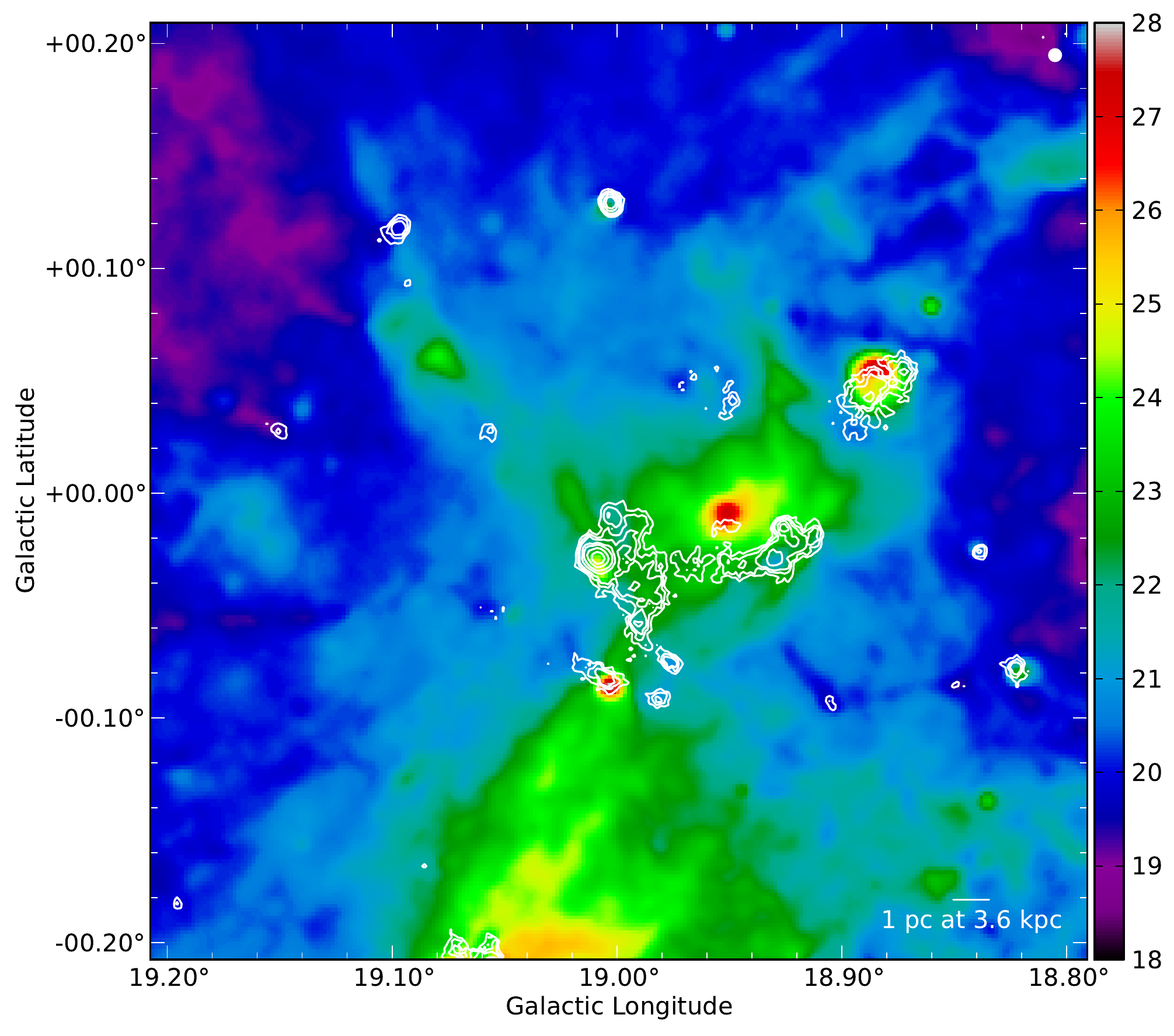}
  \includegraphics[width=0.5\textwidth]{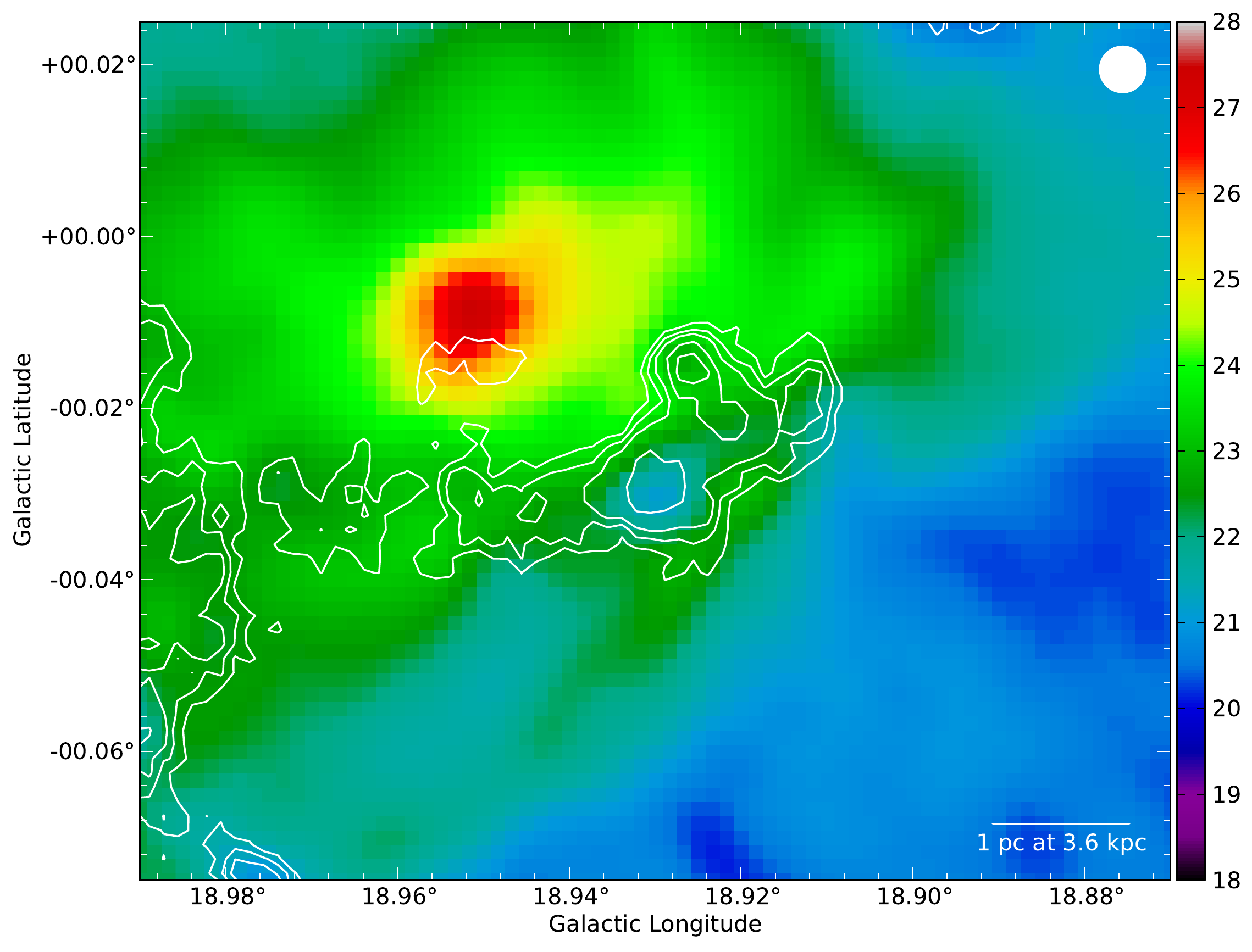}
  \caption{Color coded temperature maps in [K] of G18.93, shown by the white
    contours from ATLASGAL (at 0.3\,Jy, 0.4\,Jy, 0.5\,Jy, 0.7\,Jy, 0.9\,Jy,
    1.3\,Jy, 1.8\,Jy, and 2.5\,Jy). The beam size is indicated in the top
    right corner. While the top panel presents the large-scale environment,
    the bottom panel zooms into our central IRDC region.}
  \label{fig:temp_map}
\end{figure}
The dust temperature is not only a necessary quantity for calculating column
density and mass estimates, but it is an important physical parameter. As
described in Sec. \ref{sec:higal_description}, we can use the HiGal/HERSCHEL
survey together with ATLASGAL data to estimate the dust temperature. Smoothing
all data to the same resolution, we used frequency-dependent, optically thin
dust emission models from \citet{Ossenkopf1994} to fit every pixel with a single
temperature Planck function. (See also \citealt{Ragan2012}.) A common problem in the context of HERSCHEL PACS/SPIRE data is the unknown background contribution. Several efforts have been made to solve that problem, e.g. \citet{Stutz2010, Battersby2011}. In the context of determining the temperature we tested different background levels based on Gaussian fits to the noise distribution in regions with no or little signal. While the absolute temperatures differ up to 10\%, the relative temperature distribution is very similar. Therefore we refrain from subtracting any background and focus on the relative changes. 

Furthermore, we have compared temperature maps using both HiGal and ATLASGAL,
smoothed to a common resolution of 37\arcsec, to temperature maps with only three
bands, 70\micron, 160\micron, and 250\micron at a resolution of
19\arcsec. Although it is ambitious to fit a curve to only three data points,
the peak of the SED is covered and can be reconstructed. The ATLASGAL data has
been omitted since the missing background subtraction implies problems with the
calibration and worsens the reconstruction of the peak. Comparing the
temperatures of both SED fits on regions where differences due to the beam
size are negligible, we find very good agreements. We want to point out
  that omitting the longer wavelength data biases the absolute results towards higher
  temperatures. However, the relative distribution in the temperature maps are
  preserved. Therefore, Fig. \ref{fig:temp_map} shows the temperature maps at the better resolution of 19\arcsec. To have more reliable absolute temperatures for deriving column densities and masses, we use the temperatures derived on all HiGal plus the ATLASGAL data.

As displayed in the top panel of Fig. \ref{fig:temp_map}, the dust around the
IRAS source is hottest and produces a large scale temperature
gradient. Nevertheless, all ATLASGAL peaks are colder than their neighborhood,
and the temperature dip towards G18.93/m is the largest.

\subsubsection{G18.93/m: a starless clump}
\label{starless_clump}
Despite its prominence as absorption feature at SPITZER wavelengths,
\citet{Tackenberg2012} did not list G18.93/m as a starless clump. Because of a
peak in the extended 24\micron emission within the clump boundaries, the clump
had been rejected for consistency. In this detailed study however, we refrain from
such rigorous methods and allow an individual inspection of the clump. Using
the GLIMPSE color criteria given in \citet{Gutermuth2008}, none of the
cataloged near- and mid-IR sources within the clump is classified as young
stellar object (class 0/I). Using similar GLIMPSE color criteria,
\citet{Robitaille2008} did not identify young stellar objects
projected onto the clump either. For even younger sources not yet
visible at near-IR wavelength, Fig. \ref{fig:G1893_zoom} shows that
there is no point source longwards of 8\micron close to the ATLASGAL
peak of G18.93/m. To further quantify our sensitivity we
estimated the luminosity of objects that still could be embedded in
the dust. Using the same assumptions as explained in
Sec. \ref{sec:dust_temp}, we here fitted two blackbody functions to the point source detection limits of the 24\micron and
70\micron images of 2\,mJy for MIPSGAL \citep{Carey2009} and
$\sim$\,10\,mJy for the 70\micron HiGal field together with the flux estimates for the absorption peak of G18.93/m at 160\micron
and 250\micron, 52.0\,Jy and 45.1\,Jy. The warmer component then has
only 0.1\,L$_\odot$, corresponding to a low-mass star of
0.14\,M$_{\odot}$. This can be considered as an upper limit for any potential
embedded and undetected source. 

Additionally indicative of true starless clumps is the absence of SiO emission. Since SiO traces outflows it is commonly used to differentiate between starless clumps and star forming clumps \citep{Motte2007, Russeil2010}. For the entire G18.93 complex we find no SiO down to column densities of $\sim$ 8\e{11}\,cm$^{-2}$. 


In order to understand whether G18.93/m is pre-stellar or only a transient object, we compare its virial mass to its dust mass. While pre-stellar
clumps are gravitationally bound and therefore will eventually form stars in the future, transient clumps are not gravitationally bound. Within such
objects, smaller (here unresolved) fragments can still collapse, but the clumps itself will drift apart. As we will discuss in
Sec. \ref{sec:spec_imprints}, we find a double peaked \hdzco profile at the ATLASGAL peak position of G18.93/m. As shown in
Fig. \ref{fig:spec_imprints}, the line width of the component we attribute to the IRDC is $\delta$v\,=\,2.1\,km/s. (For details see
Sec. \ref{sec:spec_imprints}.) To calculate the virial mass we use the equation given in
\citet{MacLaren1988}, M$_{vir}$\,=\,k\,R\,$\delta$v$^2$. For the source
radius R we use 0.5\,pc as given in Table \ref{tab:clump_mass}, and the geometrical parameter k depends on the density structure of the clump and is k\,=\,190 for $\rho$ $\propto$ 1/r, k\,=\,126 for $\rho$ $\propto$ 1/r$^2$. The mass then becomes M$_{vir}$\,=\,420\msun for a geometrical parameter k\,=\,190 or $\rho$
$\propto$ 1/r, and M$_{vir}$\,=\,280\msun with k\,=\,126 or
$\rho$ $\propto$ 1/r$^2$. \citet{Beuther2002a} and \citet{Hatchell2003} find
typical density distributions in sites of massive star formation of $\rho$
$\propto$ r$^{\alpha}$ with $\alpha$\,$\sim$\,-1.6, in between both parameters. Doing a
linear interpolation, the suggested virial mass becomes 360\msun. 
Considering all uncertainties, a comparison to the estimated dust mass for G18.93/m of 280\msun does
not allow a conclusive distinction whether clump G18.93/m is gravitationally
bound or not. However, as discussed in Sec. \ref{sec:dust_temp} the dust
temperatures derived from SED fitting to HERSCHEL data are very uncertain due
to missing background levels. Comparing gas temperatures measured with NH$_3$
observations \citep{Pillai2006} to dust temperatures from HERSCHEL observations
\citep{Henning2010}, we find that despite the efficient dust cooling the
ammonia temperatures are usually lower. Thus, the given temperatures are only
upper limits and the clump masses may be higher. Therefore, the virial
analysis is consistent with G18.93/m being gravitationally bound, and pre-stellar.

\subsection{The bubble}
\begin{figure}[tbp]
  \includegraphics[width=0.5\textwidth]{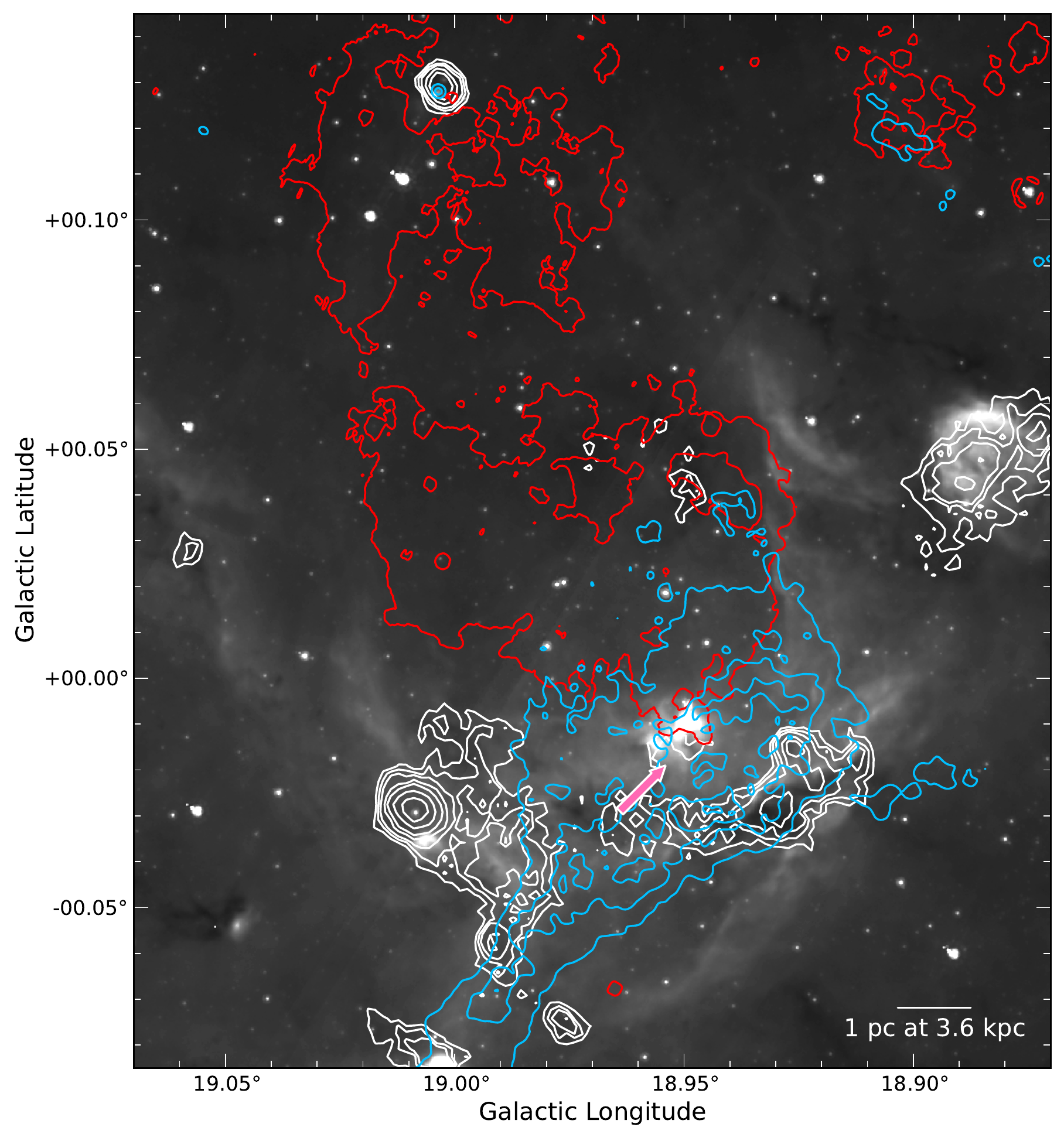}
  \caption{The ionized gas within and around the
    bubble. On top of the GLIMPSE 8\micron image, the red contours show the H$_{\alpha}$
    emission from the SuperCOSMOS survey, the blue contours represent the
    MAGPIS 20\,cm emission, and white contours show the ATLASGAL emission (at 0.3\,Jy, 0.4\,Jy, 0.5\,Jy, 0.7\,Jy, 0.9\,Jy, 1.3\,Jy, 1.8\,Jy, and 2.5\,Jy). The contour levels for the MAGPIS data are 0.002\,Jy/beam, 0.003\,Jy/beam, 0.004\,Jy/beam, and 0.005\,Jy/beam. The arrow indicates the peak position of the \hii region identified by \citet{Lockman1989}.}
  \label{fig:all_ionized}
\end{figure}
According to \citet{Simpson2012}, the bubble connected to the filament and described in
Sec. \ref{sec:irdc_bubble}, or MWP1G018980+000304, has an effective radius
of 3.44\arcmin. However, as mentioned before, its appearance is more similar
to a 'champagne flow' (cf. Fig. \ref{fig:glimpse_general}). Since the
expansion velocity of an \hii region depends on the surrounding density, an
inhomogeneous medium leads to asymmetric bubbles. If the ionizing gas reaches
the edge of its parent molecular cloud or encounters dense gas, it will fan
out towards the empty space \citep{Stahler2005}. Therefore, the effective radius and the major- and minor-axis can describe the current appearance of a bubble, but cannot describe the expansion history.

Thus we have to ask: What is the origin of MWP1G018980+00304?
Galactic bubbles are usually produced by OB stars or clusters, driving an expanding H{\sc ii} region, or by stellar winds and radiation pressure of late B-type stars \citep{Elmegreen1977, Churchwell2006, Deharveng2010, Simpson2012}. For both processes, the exciting source is not necessarily in the middle of the bubble.

Often, (ultra) compact H{\sc ii} regions (UCH{\sc ii}s) are found connected to
the bubble driving source, but e.g. CORNISH \citep{Purcell2008} does not find
an UCH{\sc ii} region inside the bubble. However, despite its high
  sensitivity at 6\,cm of better than 2\,mJy, due to interferometry techniques,
  CORNISH is not sensitive to
  sources larger than 12\arcsec \citep{Hoare2012}. Searching for more
  extended structure, \citet{Lockman1989} conducted a H$_{\alpha}$ radio recombination line single dish survey at 3\,cm and
detected an H{\sc ii} region at l=18.954, b=-0.019. As shown in
Fig. \ref{fig:all_ionized}, this is right between the IRAS source and G18.93,
but measured with a beam of 3\arcmin. Although they measure a source velocity
of v$_{LSR}$\,=\,(52.3$\pm$1.7)\,km/s, we believe that the bubble and
the H{\sc ii} gas are spatially connected. Indeed, velocity offsets between the ionized gas and CO are common \citep{Blitz1982,Fich1982}.
For the same H{\sc ii} region \citet{Kuchar1997} measure the flux at 4.85\,GHz
with a beam width of 4.2\arcmin. For an estimated source diameter of
7.0\arcmin they find 2146\,mJy. Using the formula given in \citet{Kurtz1994}
and assuming a correction factor for the optical depth of a\,=\,0.9922
(corresponding to a frequency $\nu$\,=\,5\,GHz, and temperature T\,=9000K as
given in \citealt{Mezger1967}), this flux translates to the number of Lyman continuum photons n$_{Ly}$\,=\,log(N$_{Ly}$)\,=\,48.4\,s$^{-1}$.

The blue contours in Fig. \ref{fig:all_ionized} show the MAGPIS GPS data at
20\,cm. The emission peak of the high resolution data at 20\,cm agrees well
with the extrapolated peak position from \citet{Lockman1989} in the sense
  that it also lies
between the IRAS source and G18.93. However, the GPS data shows 20\,cm
emission towards the IRDC and beyond. The higher latitude flux within the contour
level 3\e{-3}\,Jy/beam, without the tail towards lower latitudes, becomes 1.5\,Jy. With a\,=\,0.9951
at 9000\,K \citep{Mezger1967}, that converts to a Lyman continuum flux of
n$_{Ly}$\,=\,48.2\,s$^{-1}$. A slightly larger contour drawn along the visual
extent of the emission provides a flux of 2.5Jy, thus 2/3rd bigger. However, with n$_{Ly}$\,=\,48.4\,s$^{-1}$ the logarithm does not change as much and the given differences can be considered as uncertainty in our measurements. Therefore, the results of \citet{Kuchar1997} and our measurements agree within the uncertainties. 

As mentioned in Sec. \ref{sec:intro_ionized}, the synchrotron radiation of
cosmic ray electrons can also contribute to the cm continuum emission. However,
the spectral index $\alpha$ (with S$_{\nu}$\,$\propto$\,$\nu^{-\alpha}$) of
the thermal free-free emission is negative in this regime, while the spectral index of the synchrotron
radiation is positive. Therefore, we can use the two independent measurements
of the continuum emission to at least determine the sign of the spectral
index. The slope between the 20\,cm, or 1.5\,GHz, data point and
the 6\,cm, or 4.85\,GHz, data point is positive, suggesting a negative
spectral index, which is in agreement with
the expected spectral index for thermal free-free emission. Therefore, the
emission is dominated by free-free radiation and we can
neglect any synchrotron contribution. Nevertheless, with only two data points
we cannot distinguish whether the assumption of optically thin emission is
correct. Therefore, the given Lyman continuum fluxes are lower limits.

While the cm continuum is more prominent towards the dust continuum emission,
large parts of the bubble are filled by SuperCOSMOS H$_{\alpha}$ emission, see
Fig. \ref{fig:all_ionized}. As mentioned in Sec. \ref{sec:intro_ionized}, H$_{\alpha}$ is a
complementary tracer of ionized gas. The difference in the spatial
distribution can be explained by the optical depth. Since the optical
H$_{\alpha}$ line is susceptible to extinction, H$_{\alpha}$ can only be
detected in regions of low visual extinction. However, the near-IR extinction
map shows that the visual extinction towards the
ATLASGAL emission peaks are larger than A$_V$\,$=$\,25\,mag.

Using the averaged number of Lyman photons, n$_{Ly}$\,=\,48.3\,s$^{-1}$ to determine the spectral type, the exciting source needs to be at least a main sequence star of type O8.5 \citep{Panagia1973, Martins2005}. The extended appearance (and the connected non-detection within CORNISH) and the morphology of the H{\sc ii} suggests that the ionizing source is not extremely young. Nevertheless, the diameter of the bubble does not allow for extremely old bubble-driving sources. 


From the mid infrared data, a good candidate for the ionizing source is
IRAS 18227-1227, marked in Fig. \ref{fig:glimpse_general}. It is very bright
at 24\micron, but by 3.6\micron it is no longer the brightest source
within its neighborhood. At higher latitudes, $\sim$ 16\arcsec above IRAS
  18227-1227, a near IR source, hardly visible at 24\micron, is another
potential candidate to drive the bubble. Although the additional source
appears blue in the four IRAC bands from GLIMPSE, the color criteria
given in \citet{Gutermuth2008} identify it as young stellar object. At the
even shorter wavelengths of the 2MASS and UKIDSS near IR JHK surveys, the
second source has a bright neighbor. Nevertheless, its missing detection in
the GLIMPSE survey and its blue appearance at NIR wavelength indicates that it
is evolved. However, the near infrared colors do not allow a classification or
mass determination. 
While for one its classification as young stellar object (class I) hints to non photospheric emission at near infrared wavelength, the second source shows a significant color excess in the JHK color-color diagram indicating non photospheric emission as well. Therefore we can not identify the exact ionizing source with the current data. Nevertheless, as shown before from cm free-free emission, it has to be of spectral type O8.5 or earlier. 

Another result one can determine from the different tracers of the ionized gas
are the densities above and below the ionizing source. While we have
material that is optically thick to visible light suggesting dense gas towards the
dust continuum emission, we have optically thin low density material within
the bubble and above it. That explains the position of the ionizing source and the 'champagne flow'. 

\subsection{The photon dominated region: a layered structure}
\label{sec:pdr}
What is the observable impact of the \hii region on the dense filament, described in Sec. \ref{sec:results}?

During the evolution of an \hii region the temperature difference between the hot ionized gas and the cold environment drives a supersonic shock front beyond the ionization front. In addition, at the interface between an H{\sc ii} region and the neighboring gas a photon dominated region builds up. While hydrogen ionizing
radiation with energies above 13.6\,eV produces the H{\sc ii}, beyond the
ionization front other molecules with lower ionization energies may still be
ionized. However, connected to the photon dominated region one expects a layered structure of H{\sc ii} - H{\sc i} - H$_2$ \citep{Hollenbach1997}. For our bubble we can directly observe the layers between the presumably ionizing source and the dense gas of G18.93/m.

While the ionized gas and its structure is directly traced by the cm free-free emission, the ATLASGAL emission connected to G18.93/m represents the cold dust, which is embedded in dense H$_2$. In order to trace the morphology of the atomic hydrogen we employ the VGPS \hi survey.

(Hot) H{\sc i} is so abundant in the Galaxy, that along the Galactic plane,
background H{\sc i} emission is present at all velocities. Therefore, cold
sources in the foreground appear as absorption features superimposed on the
background \hi emission. However, since the H{\sc i} distribution and temperature vary strongly within the Galactic disk, the observed background varies strongly as well. In addition, individual H{\sc i} clouds often have sufficiently high column densities to become self absorbing. This complicates the identification of complexes in the \hi data. 

\begin{figure}
  \includegraphics[angle=-90,width=0.5\textwidth]{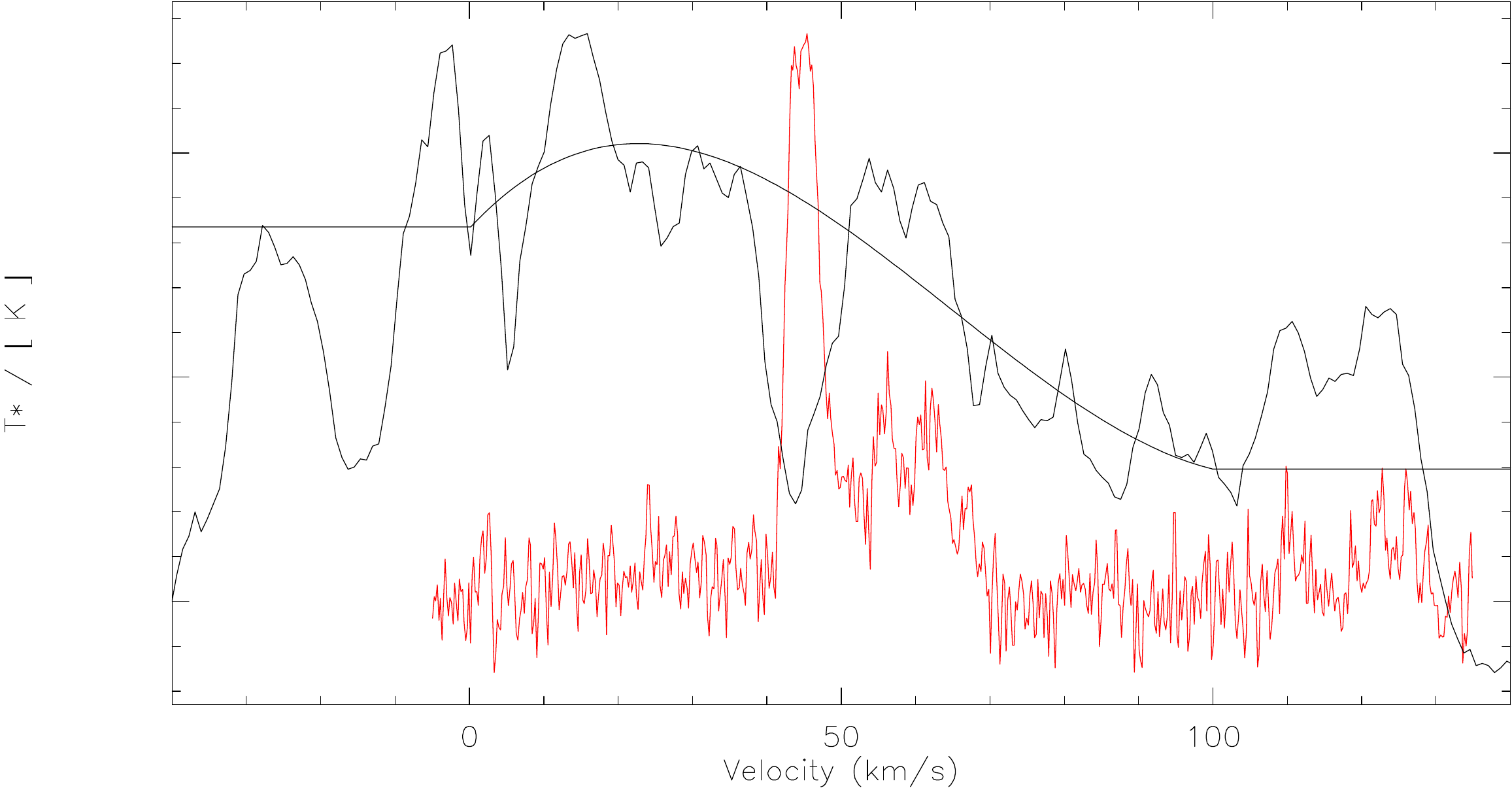}
  \caption{HI VGPS spectrum of G18.93/m in black with the GRS \dzco spectrum
    in red on top. The strong absorption feature of G18.93/m at
    $\sim$\,45\,km/s in the \hi spectrum agrees very well with the \dzco
    signature. The smooth line indicates the 3rd order fit to the \hi spectrum
    that has been used to subtract the 'continuum'. (For details see text.)}
  \label{fig:hi_emission}
\end{figure}
However, as shown in Fig. \ref{fig:hi_emission}, the absorption dip for the dense gas around G18.93 is very prominent. In order to study the distribution of the atomic hydrogen, we quantified the absorption feature by the following process: 
{\bf (1)} We fit the background in the spectra with a third order polynomial
omitting the velocity range from 40\,km/s to 50\,km/s. This determines the
large scale background variations, but preserves the signals on small
scales. {\bf (2)} We subtract the fitted function as a baseline. {\bf (3)} We
invert the spectra and fit the now emission peak with a Gaussian. The
integrated area of the
Gaussian for a constant temperature is proportional to the column
density. (NOTE: We stress that the authors do not want to convert the so
quantified measure of \hi into physical meaningful units, but only take it as
relative value.) Doing this procedure on all \hi spectra around G18.93/m, we
obtain the qualitative distribution of the atomic hydrogen. \\
A problem with this method arises from the free-free continuum. Due to the
additional background emission towards the \hii region, the absorption dip
becomes stronger for the same \hi column densities. Therefore, the
interpretation of the \hi distribution needs to be considered with
caution. However, since the peak of the \hi and \hii emission do not agree, the
suggested \hi distribution seems to be dominated by the \hi column density.

\begin{figure}
  \includegraphics[width=0.5\textwidth]{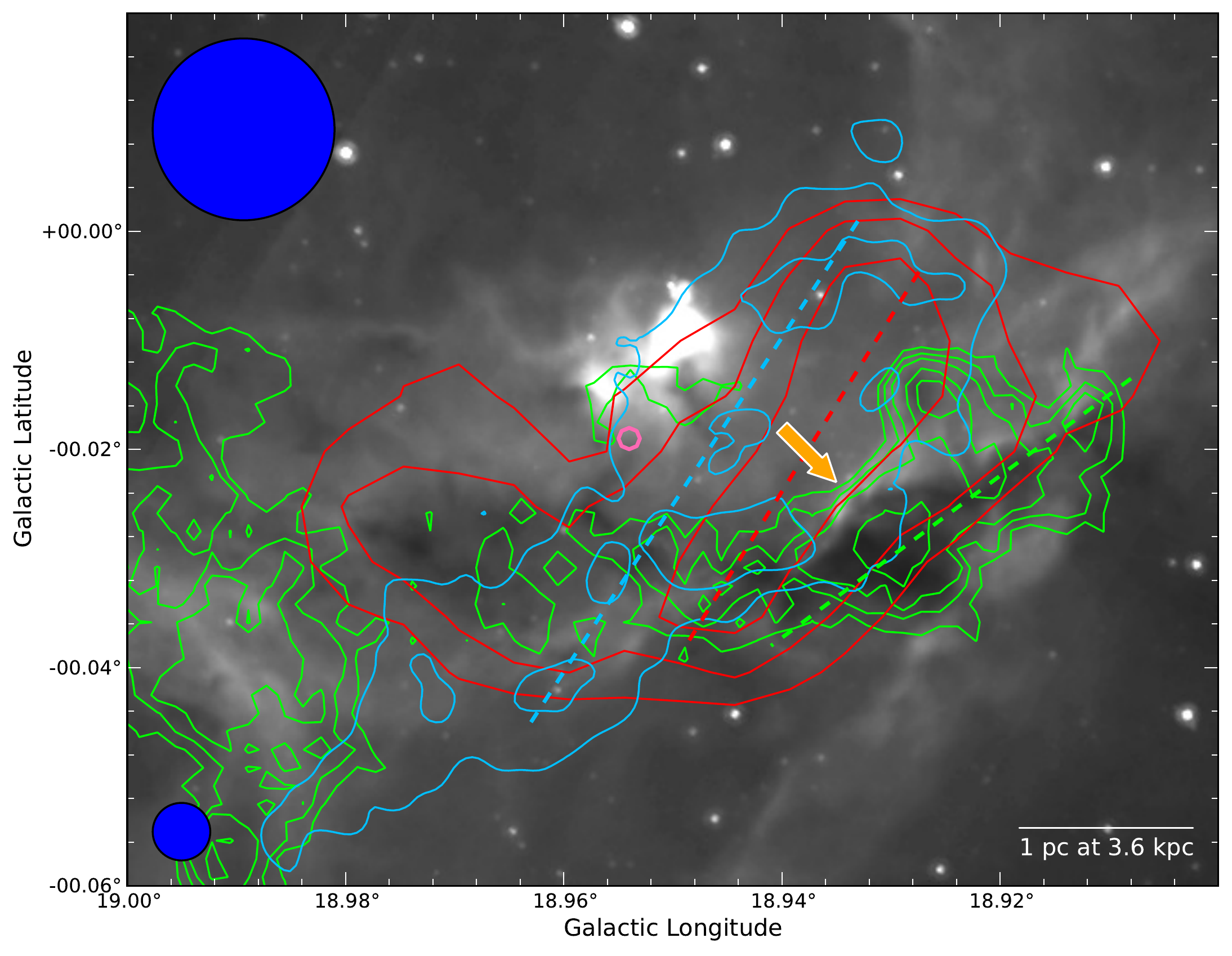}
  \caption{A layered structure of the hydrogen phase. On top of the 8\micron GLIMPSE image in gray, the blue contours represent the
    MAGPIS GPS 20\,cm data, the red contours show a measure of the atomic
      hydrogen at velocities of G18.93 from the VGPS 21\,cm line (for details see the text), 
    and the green contours are the cold dust from the ATLASGAL
    survey, hence the molecular hydrogen. The dashed lines are to guide the
    eye and indicate the peak positions of the different data. The orange arrow
    points to 8\micron excess emission at the edge of the dense gas. The top
    left circle indicates the beam size of the VGPS data, while the bottom
    circle shows the beam of ATLASGAL. In this figure, MAGPIS is smoothed to
    the resolution of ATLASGAL. The contour levels for the GPS 20\,cm data are
    2\,mJy/beam, 3\,mJy/beam, 4\,mJy/beam, and 5\,mJy/beam, and
    for the ATLASGAL 870\micron emission 0.3\,Jy, 0.4\,Jy, 0.5\,Jy, 0.7\,Jy, 0.9\,Jy, 1.3\,Jy, 1.8\,Jy, and 2.5\,Jy.}
  \label{fig:layers}
\end{figure}

Figure \ref{fig:layers} shows three physical states of hydrogen. Between the ionizing source and G18.93 a layered structure is visible, with first the ionized gas, then the atomic hydrogen, and then the molecular gas within the IRDC. However, one should keep in mind that the beam of the \hi data is on the same order as the spacing between the different peaks.

\section{Imprints of triggering? Discussing the interaction between G18.93/m and the expanding \hii region}
\label{sec:imprints}
So far, most studies of triggered star formation (e.g. \citealt{Deharveng2003},
\citealt{Zavagno2006})
have searched for young stellar objects connected to bubbles and \hii regions. To
understand the statistical significance of triggered star formation, current
and future studies make use of the rising number of Galactic plane
surveys \citep{Zavagno2010, Deharveng2010, Kendrew2012}. Although it has
been shown in the past that not all stars on the border of \hii regions are
necessarily formed by triggering, it becomes clear that triggering might have a
significant effect on star formation.

As a next step towards a better understanding of triggered star formation we
need to observationally identify the mechanisms that govern the interaction
between expanding \hii regions and the molecular gas. In this context,
detailed studies of individual regions are needed to reveal signatures of
triggering before star formation sets in (e.g. \citealt{Bieging2009}).

In the following Section we will explore whether we can see any imprints of
the \hii region on the starless clump G18.93/m.

\subsection{Comparison of G18.93/m with typical high-mass starless clumps}
\label{sec:aver_starlessclump}
As discussed in Sec. \ref{starless_clump}, G18.93/m is a starless clump with
no IR emission up to 160\micron. While the virial analysis does not allow a
  firm conclusion as to whether it is bound or not, but if it is bound it would be a proto-type pre-stellar clump. In terms of mass and size it does not stand
out. If we compare it to the starless clumps found in \citet{Tackenberg2012},
its clump mass of 280\msun is comparable to the average mass of 315\msun they
find. Note that with the lower dust opacity of $\kappa$\,=\,0.77\,cm$^2$ g$^{-1}$ used in
\citet{Tackenberg2012} as well as the low temperature of 15K they assume for
starless clumps, the clump mass becomes 900\msun. Nevertheless, for peculiar
sources such as G18.93/m, the elevated dust temperature needs to be taken into
account. 

If we assume a spherical clump we can calculate the average volume density of
G18.93/m to be 1\e{4}\,cm$^{-3}$. This is significantly smaller than the
5.0\e{4}\,cm$^{-3}$ for the sample of starless clumps presented in
\citet{Tackenberg2012}. However, from the detection of \hdzco with a critical
density of $\sim$\,1.8\e{5}\,cm$^{-3}$, we know that at least in the central
regions the density needs to be higher. Nevertheless, since the average
densities of the starless clumps have been calculated with the same
assumptions, the results should be comparable. For the formation of high-mass
stars, \citet{Krumholz2008} require a peak column density of
3\e{23}\,cm$^{-2}$. With a peak column density of
4\e{22}\,cm$^{-2}$, G18.93/m does not meet their requirements. However, the
measured column densities are beam averaged values, true peak column densities
are expected to be higher \citep{Vasyunina2009}. 
Although the expanding \hii region might have influenced the starless clump G18.93/m, it does not differ from other starless clumps. 

\subsection{The photon dominated region: a layered structure}
\label{sec:discuss_pdr}
While the general distribution of the atomic hydrogen follows the ATLASGAL
emission, the peak of \hi is elongated towards the gap between the ionizing
source and the molecular hydrogen, shown by Fig. \ref{fig:layers}. Towards
G18.93/5 (cf. Fig. \ref{fig:glimpse_general}), the 20\,cm free-free
emission is no longer parallel to the cold dust, but crosses the IRDC.
However, both extinction and dust emission are significantly lower at the
intersection. That implies that there is less dense gas which allows the
ionized gas to more easily escape beyond G18.93/5; the ionizing radiation is simply no longer blocked across the full height of the filament.

In the context of star formation, \citet{Glover2012} have shown that the
composition of the gas has hardly any influence on the star formation
efficiency. They suggest that molecular gas is not a prerequisite for the
  formation of cold clumps/cores, but that the high (column) densities required also
  prohibit the destruction of molecules. Hence molecular gas is not a
requirement, but form simultaneously. Therefore, the layered structure
does not affect the star formation within G18.93. Instead, shielding of the
radiation field is most important and the layered structure proves the effectiveness of the shielding. 


\subsection{The temperature distribution}
\label{sec:discuss_temp}
As described in Sec. \ref{sec:dust_temp}, the hot ionizing source as well as the IRAS source heat up their environment and produce a temperature gradient across G18.93 (cf. Fig. \ref{fig:temp_map}). On large scales shown by the top panel of Fig. \ref{fig:temp_map}, beyond the heating from the ionizing source the fitted dust temperature drops to values between 19\,K and 20\,K, typical for the ISM \citep{Reach1995}. Within, the temperatures towards the continuum peaks stand out against their direct vicinity. However, all dense clumps are warmer than the general dust/ISM temperature. 

\begin{figure*}
  \includegraphics[width=\textwidth]{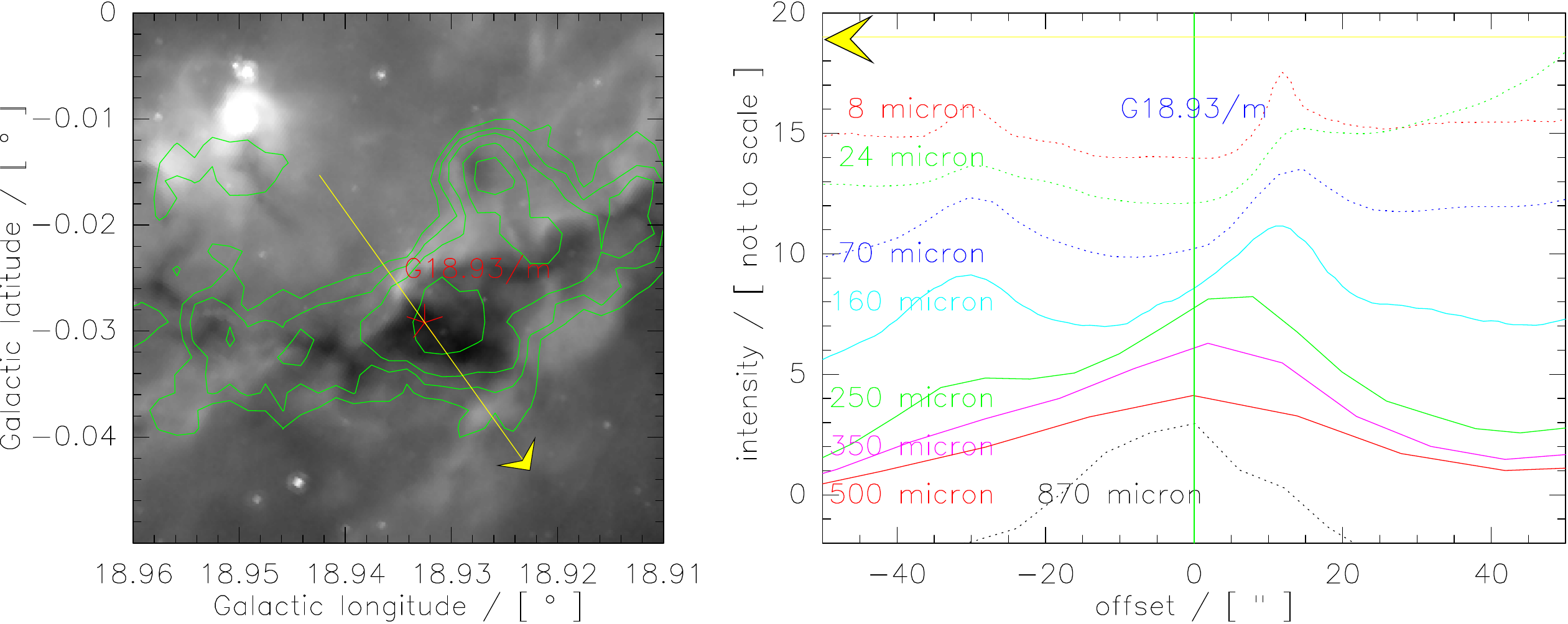}
  \caption{Profiles of mid IR data along the cut
    indicated by the yellow arrow in the left panel. While the center
  of the cut in the right panel is G18.93/m,
    its upper end is chosen to be towards the potential ionizing
    sources. The different colored profiles in the right panel are of increasing wavelength, starting with
    8\micron at the top going to 870\micron at the bottom. The intensities are not to scale, but adopted to fit in
    the panel. The green vertical line indicates the position of the
    ATLASGAL continuum peak, the yellow arrow corresponds in direction
    and length to the arrow in the left panel, in which the background image
    is a GLIMPSE 8\micron image with ATLASGAL contours (at 0.3\,Jy, 0.4\,Jy, 0.5\,Jy, 0.7\,Jy, 0.9\,Jy, 1.3\,Jy, 1.8\,Jy, and 2.5\,Jy) on top.}
  \label{fig:higal_profiles}
\end{figure*}
A different visualization of the temperature gradient becomes visible if we
compare the mid-IR wavelengths. Figure \ref{fig:higal_profiles} shows a cut
through G18.93/m for different wavelengths, ranging from 8\micron to
870\micron. While the longest wavelengths trace the column density peak of the
cold dust, the SED peak moves along the temperature gradient towards the
heating source. Therefore, the shorter wavelengths are offset towards the
heating source as well and we can trace its transition directly if we employ
the full resolution at all wavelengths. Only the 8\micron band is not
consistent with that picture. Since it is dominated by PAH emission, its peak
is influenced by the ionizing radiation.

IRDC temperatures elevated above the general ISM temperature of $\sim$ 18\,K
are different from typical regions of both low- and high-mass star formation
(\citealt{Peretto2010a}, \citealt{Battersby2011}, Nielbock et al., submitted,
Launhardt et al., in prep.) in which their temperatures (or the ones of dense
cores) drop to temperatures below the ISM value. In contrast to the above studies, \citet{Beuther2012} find values for IRDC\,18454 similar to what we find. In IRDC\,18454, the mini starburst cluster W43 raises the general temperature of the IRDC complex to similar temperatures as we find for G18.93. 
\citet{Beuther2012} point out that the elevation of the dust temperature
raises the Jeans length/mass, and therefore speculate that high-mass star
formation may be favored. Hence, it could be that it is not the expansion of the
\hii region and the connected shock front that promotes the formation of OB
associations on the rims of bubbles, but the effects of heating on the environment.

Therefore we conclude that the bubble does have an effect on the temperature of the IRDC, but we cannot distinguish whether this is only due to the heating source or also because of the shock front.

\subsection{Does the \hii region change the shape of the IRDC?}
\begin{figure*}
  \includegraphics[angle=-90,width=\textwidth]{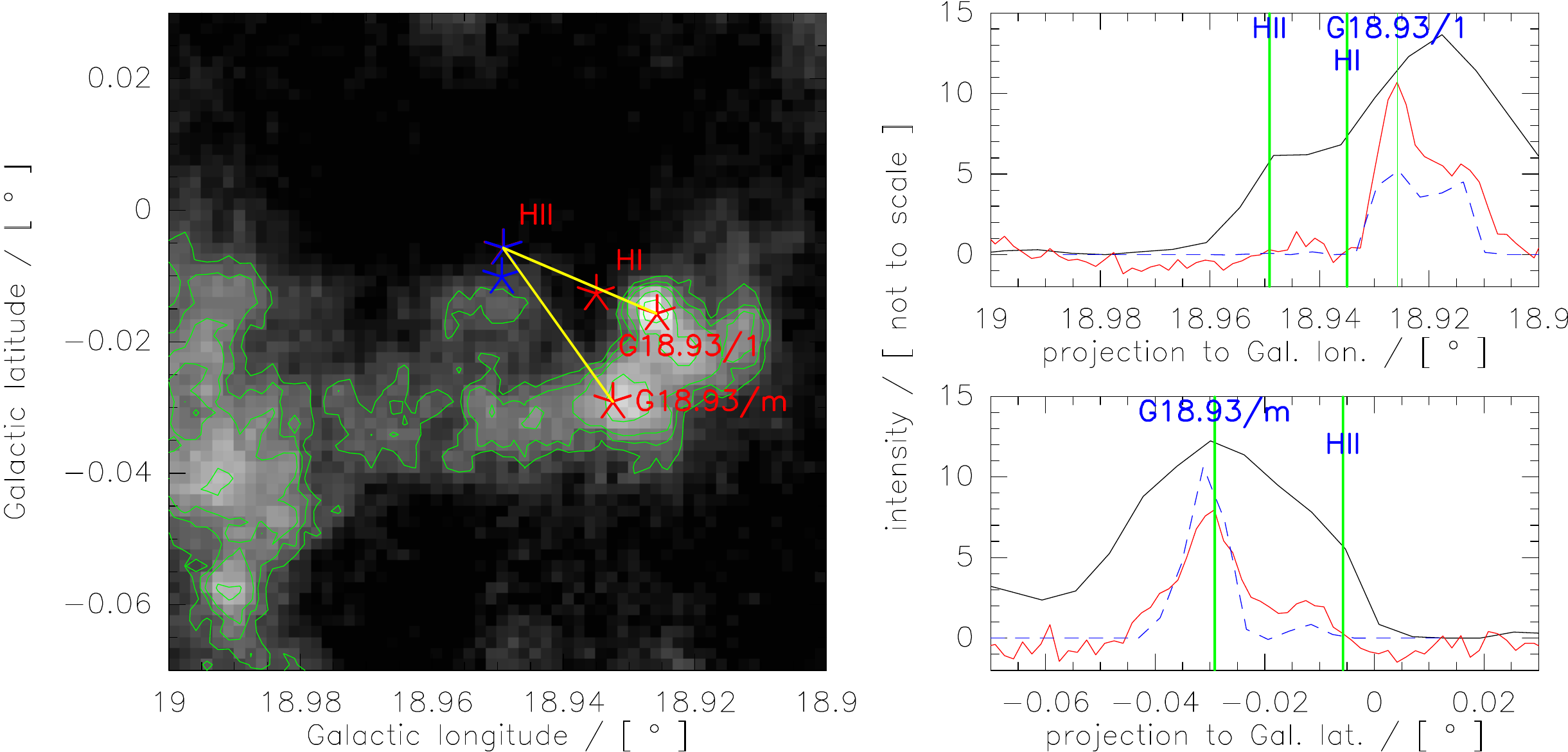}
  \caption{Profile of \dzco (black), ATLASGAL (red),
    and \hdzco (dashed blue) along the two yellow lines shown on top of
    the 870\micron map in the left plot. The position of the peak of \hii
    emission is indicated within both profiles (right) and the map
      (left). The left panel also gives the two main IRDC peaks. The blue
    asterisks indicate the position of the IRAS source (lower position) and
    UKIDSS source (higher position). The green contours from ATLASGAL are at 0.3\,Jy, 0.4\,Jy, 0.5\,Jy, 0.7\,Jy, 0.9\,Jy, 1.3\,Jy, 1.8\,Jy, and 2.5\,Jy.}
  \label{fig:density_profiles}
\end{figure*}
\citet{Elmegreen1998} explains the formation of fingers (or elephant trunk like structures) as a shock wave runs over density enhancements and this has been directly observed (e.g. \citealt{Motte2010}). For the interface between G18.93 and the neighboring bubble no such structures are visible in Fig. \ref{fig:glimpse_general} nor at other wavelengths available. 

In the picture of the 'collect and collapse' model the \hii region pushes material along. In the vicinity of a pre-existing IRDC, the shock wave mainly penetrates one side of the IRDC, the one facing the \hii region. Thus the IRDC should become asymmetric, with a steeper density profile towards the bubble. Such an asymmetry should be reflected in the projected profiles of the IRDC. 

Figure \ref{fig:density_profiles} shows profiles of various gas tracers along
two lines. While the profile through clump G18.93/1 shows a second peak along
the profile and is therefore unsuited for comparing the shape of the profile,
the cut through G18.93/m seems to be well suited. Figure
\ref{fig:density_profiles} shows that \dzco is more of an envelope around the
entire complex and the \hii peak is enclosed within the low density
gas. Different from that, the dense gas peak from ATLASGAL data is clearly narrower. However, fitting both sides of the sub-mm continuum intensity distribution of G18.93/m with an exponential function independently, we find no difference between both sides of the profile. At the resolution of the IRAM \hdzco maps we do not have sufficient data points to do statistically meaningful fits to both slopes, but on the sparse data available, again we find no difference. 

Therefore we conclude that at least for G18.93/m the bubble does not seem to
influence the density profile of the IRDC.

\subsection{Imprints of the shock in the dense gas?}
\label{sec:spec_imprints}
\begin{figure*}
  \includegraphics[width=\textwidth]{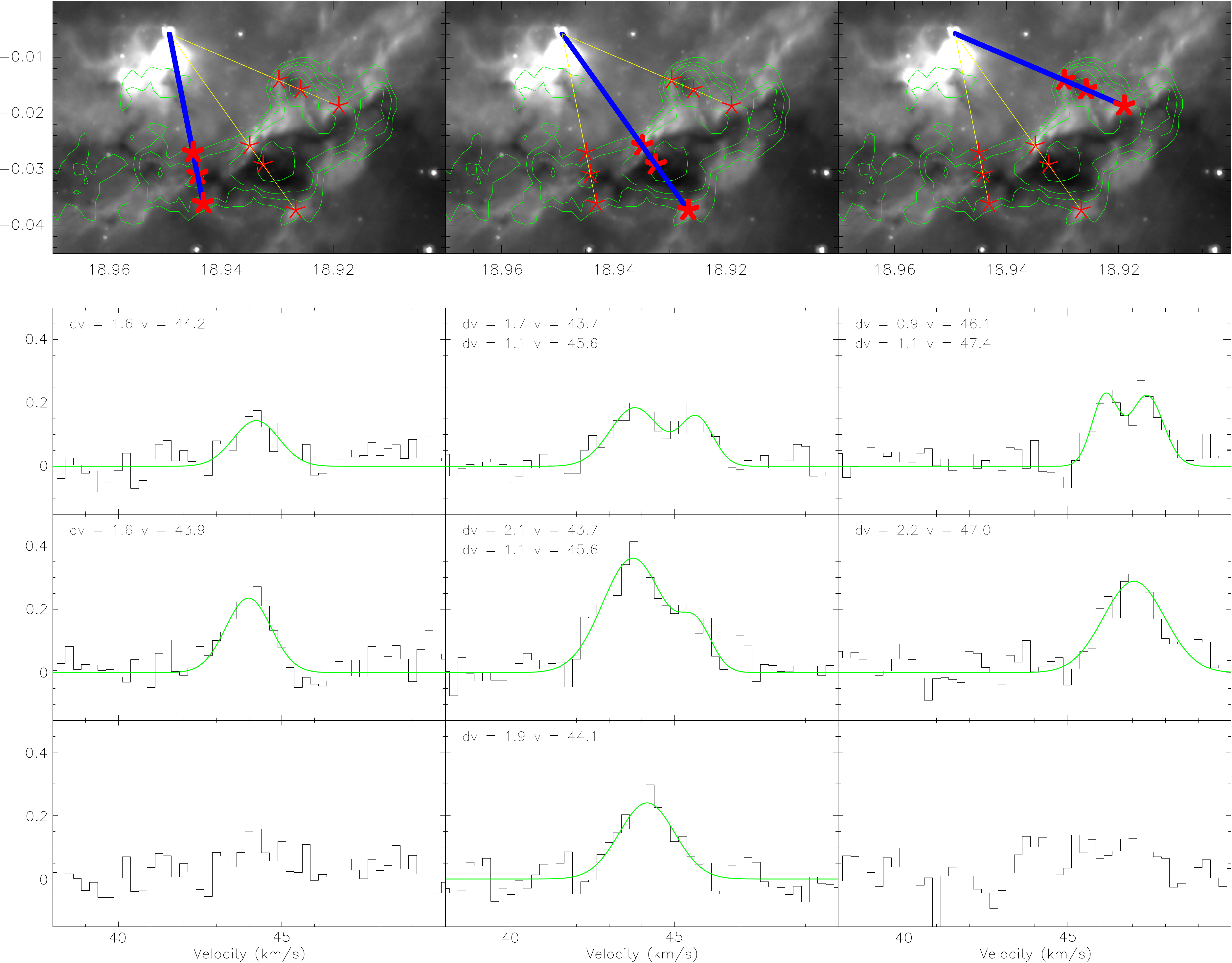}
  \caption{\hdzco spectra at positions indicated by bold asterisks in the top
    panel. The positions are chosen to lie along the connection between the
    UKIDSS near-IR source and the ATLASGAL peak position, with the first spectra at the peak of the GLIMPSE 8\micron emission. From left to right the clumps are G18.93/1, G18.93/m, and G18.93/4. The upper spectra are at positions closer to the ionizing source. Where possible we fitted one, two, or three gaussian components to the spectra. The fitted line widths and peak positions are given in the spectra.}
  \label{fig:spec_imprints}
\end{figure*}
After the fast (super sonic) initial expansion of an \hii region to the size
of a Str\"omgen sphere, the shock's velocity quickly drops below the sound
speed v$_S$ of the medium. Typical values of v$_S$ are close to 10\,km/s for
the hot ionized gas. As the shock front hits the dense gas of G18.93, the
expansion velocity drops dramatically, while the lower density gas towards
higher latitudes allows an accelerated expansion producing the 'champagne flow'. However, even the slow shock front should still leave its imprint in the dense gas. 

In order to search the potentially shocked gas for velocity peculiarities, we examined \hdzco spectra at various positions.

For three clumps, Fig. \ref{fig:spec_imprints} compares \hdzco spectra within the 8\micron rim to spectra at the clump's peak position and positions away from the \hii region. Closest to the shock front, the spectra of G18.93/m and G18.93/1, the two right panels of Fig. \ref{fig:spec_imprints}, consist clearly of two distinct velocity components of similar strength. Since \hdzco is expected to be optically thin, we can exclude self absorption. The spectra at the position of the ATLASGAL continuum peaks show two components at the same velocities. However, here one component is significantly stronger than before. At the third position, away from the \hii region, G18.93/m shows only a single component, while we do not detect signal for G18.93/1. 

A possible interpretation of those spectra could be that one component traces
the IRDC, while the second component is shock induced. In that context the
rising component towards the peak of the IRDC traces the density distribution
of the dense gas. Its distribution is almost symmetric perpendicular to the
filament. The second component is bright towards the \hii region, but due to
the IRDC's high density, the shock has not propagated beyond the peak and
therefore no shock component is visible away from the shock front. For the
spectra taken across G18.93/4, within our sensitivity limits we cannot find
similar clear imprints. Although the spectrum at its peak position has no pure
Gaussian profile and a second component is required to fit it, a similar
interpretation as above does not work. In Sec. \ref{sec:discuss_pdr}, we
argued that the layered structure breaks down clumps at higher longitudes, because of their lower density. For the same reason, the shock might have had less impact on the dense gas at the center of the filament. Therefore, the imprints in the spectra are not as clear. 

It is interesting to note that \citet{Klessen2005} finds similar imprints on
quiescent cores. For their modeling of dense cores in which the turbulence is
driven on large scales, the velocity dispersion $\sigma_{turb}$ of their
simulated cores show bow-shaped enhancements (rims) around the continuum
emission. In our picture, instead of a large scale convergent flow, the shock
introduces turbulence on large scales. The fact that \citet{Klessen2005} measure an increase of line-width instead of two separate velocity components could well be a result of their method for determining $\sigma_{turb}$.

However, the \hii region does influence the dense gas of the IRDC. The additional velocity component might give the IRDC additional turbulent support to build more massive fragments. 

Another possible imprint of a shock on dense gas could be a broadened line width. Within the limits of our \hdzco maps, we do not observe such a broadening.

\section{Conclusion}
In the middle of a $>$\,55\,pc long filament is G18.93, a prominent IRDC. While it has the
luminous protostellar object EGO G19.01-0.03 at one end, the other end is
dark up to 160\micron. Particularly interesting about G18.93 is its
environment; it is located at the projected interface of two IR
bubbles. However, it is only spatially
coincident with the bubble above the filament, while the bubble
at lower latitudes has a different
velocity. Using \hdzco as dense gas tracer we unambiguously
attribute kinematic distances to the ATLASGAL continuum emission. Although very
close in projection, the high-mass star forming region EGO G19.01-0.03 and its
cold dust is not connected to G18.93, but exists at a different distance. From a Galactic rotation curve we
determined IRDC G18.93's distance to be 3.6\,kpc. 

We used CLUMPFIND to decompose the dense gas seen by ATLASGAL into 6
clumps. Together with temperature estimates from HiGal SED
fitting, we calculated both the column density and masses of these clumps. The
most massive clump is G18.93/m with a gas mass of $\sim$ 280\msun. A
comparison of the gas mass to its virial mass of 360\msun shows that within
the uncertainties it could be gravitationally bound. To determine its evolutionary stage, we first searched the
GLIMPSE catalog for young stellar objects using IRAC color criteria, but 
no young stellar object was found within the boundaries of G18.93/m. Next we
visually inspected the longer wavelength images MIPSGAL 24\micron and
HiGal and find no point source at wavelength up to 160\micron. 

In addition, the absence of SiO
emission is strongly indicative of no star formation
activity. Therefore, we identify G18.93/m as a potential high-mass pre-stellar clump. 

Expanding the SED fitting from single positions to all pixels, we produced
temperature maps. The sources embedded in the dust connected to IRAS18227-1227 heat the dust and produce a strong gradient across the IRDC. Therefore, all IR dark continuum peaks have temperatures above the general ISM value of $\sim$ 19\,K. 

In context of the bubble we show that the 'champagne flow' structure is
produced by an expanding \hii region. From measuring the radio continuum flux
we estimate the number of Lyman continuum photons and constrain the ionizing source to be at least an O8.5 star. 

From VGPS \hi spectra we extract a measure of the \hi column density. Together with the cm continuum data and the ATLASGAL data we identify a layered structure between the ionizing source and the IRDC, from ionized through atomic to molecular hydrogen. 

Finally, we discuss the imprints of the expanding \hii region on the starless clump G18.93/m. Our main results are:
\begin{itemize}
\item {\it layered structure:} The ionizing source produces a layered structure towards the IRDC and therefore changes the composition of the hydrogen phase. 
\item {\it temperature distribution:} The IRAS source and/or the ionizing source produce a strong temperature gradient across the IRDC. Although the dust temperature at the continuum peaks is reduced compared to their surrounding, their absolute temperatures are above the ISM values. This is atypical compared to most other IRDCs. However, as discussed in \citet{Beuther2012}, the additional support raises the Jeans length which might favor high-mass star formation.
\item {\it density profiles:} In the picture of the 'collapse and collect' scenario we would expect the shock front to steepen the density profile of the dense gas. Therefore we compare the emission profile of the cold gas towards and away from G18.93/m. Fitting both wings, we find no difference. 
\item {\it imprints within the dense gas:} Looking at \hdzco spectra across the filament, we find emission that resembles the brightness of the ATLASGAL dust continuum. In addition we find a second \hdzco component, brightest within a PAH rim towards the bubble and not detected away from the bubble. We speculate that this might be a direct imprint of the shock front onto the IRDC.
\end{itemize}

Therefore, while the additional heating and the shock induced velocity
component favor high-mass star formation, we do not find evidence for
collapse triggered by the expanding \hii region.


\section{Acknowledgment}
This publication is partially based on data acquired with the Atacama
Pathfinder Experiment (APEX). APEX is a collaboration between the
Max-Planck-Institut f\"ur Radioastronomie, the European Southern Observatory,
and the Onsala Space Observatory.
This publication makes use of molecular line data from the Boston
University-FCRAO Galactic Ring Survey (GRS). The GRS is a joint
project of Boston University and Five College Radio Astronomy
Observatory, funded by the National Science Foundation under grants
AST-9800334, AST-0098562, \& AST-0100793.
The International Galactic Plane Survey is supported through a Collaborative Research Opportunities grant from the Natural Sciences and Engineering Research Council of Canada.
The National Radio Astronomy Observatory is a facility of the National Science Foundation operated under cooperative agreement by Associated Universities, Inc. 
This work is based, in part, on observations made with the Spitzer Space Telescope, which is operated by the Jet Propulsion Laboratory, California Institute of Technology under a contract with NASA.
This research has made use of the NASA/ IPAC Infrared Science Archive, which is operated by the Jet Propulsion Laboratory, California Institute of Technology, under contract with the National Aeronautics and Space Administration.
The UKIDSS project is defined in Lawrence et al 2007. UKIDSS uses the UKIRT Wide Field Camera (WFCAM; Casali et al 2007) and a photometric system described in Hewett et al 2006. The pipeline processing and science archive are described in Irwin et al (2008) and Hambly et al (2008). We have used data from the 7th data release. 
J. T. is supported by the International Max Planck Research School (IMPRS) for Astronomy and Cosmic Physics.

\bibliographystyle{aa}
\bibliography{lib}

\end{document}